\documentstyle[aps,preprint,epsf]{revtex}

\newcommand{\beq}{\begin{equation}}
\newcommand{\eeq}{\end{equation}}
\newcommand{\beqa}{\begin{eqnarray}}
\newcommand{\eeqa}{\end{eqnarray}}

\newcommand{\bfpnu}{{\bf p}_\nu}
\newcommand{\bfpi}{{\bf p}_N}
\newcommand{\bfpe}{{\bf p}_e}
\newcommand{\bfpf}{{\bf p}_{N} '}
\newcommand{\bfpnuf}{{\bf p}_\nu '}

\newcommand{\magpnu}{|{\bf p}_\nu|}
\newcommand{\magpi}{|{\bf p}_N|}
\newcommand{\magpe}{|{\bf p}_e|}
\newcommand{\magpf}{|{\bf p}_{N} '|}
\newcommand{\magpnuf}{|{\bf p}_\nu '|}

\newcommand{\rhob}{\rho_{\rm B}}

\begin{document}
\draft
\title{Neutrino opacities in neutron stars with kaon condensates}
\author{Takumi Muto\thanks{muto@pf.it-chiba.ac.jp}}	
\address{Department of Physics, Chiba Institute of Technology, 
2-1-1 Shibazono, Narashino, Chiba 275-0023, Japan}
\author{Masatomi Yasuhira}   
\address{Yukawa Institute for Theoretical Physics, Kyoto 606-8502, Japan }
\author{Toshitaka Tatsumi\thanks{tatsumi@ruby.scphys.kyoto-u.ac.jp}}   
\address{Department of Physics, Kyoto University, Kyoto 606-8502, Japan }
\author{Naoki Iwamoto\thanks{niw@eng.kagawa-u.ac.jp; niw@physics.utoledo.edu} }
\address{Department of Advanced Materials Science, Faculty of Engineering, 
 Kagawa University, Takamatsu, Kagawa 761-0396, Japan,  \\
 and Department of Physics and Astronomy, The University of Toledo, Toledo, Ohio 43606-3390, U. S. A.}

\date{\today}
\maketitle

\begin{abstract}
The neutrino mean free paths in hot neutron-star matter are obtained in the presence of  kaon condensates. The kaon-induced neutrino absorption process, which is allowed only in the presence of  kaon condensates, is considered for both nondegenerate and degenerate neutrinos. The neutrino mean free path due to this process is compared with that for the neutrino-nucleon scattering.  While the mean free path for the kaon-induced neutrino absorption process is shown to be shorter than the ordinary two-nucleon absorption process by several orders of magnitude when temperature is not very high, 
the neutrino-nucleon scattering process has still a  dominant contribution to the neutrino opacity. Thus, the kaon-induced neutrino absorption process has a minor effect on the thermal and dynamical evolution of protoneutron stars. 
\end{abstract}

\newpage
\section{Introduction}
\label{sec:intro}

Possible existence of kaon condensation in neutron star matter  has been widely investigated in connection with its implications for   astrophysical phenomena such as rapid cooling of neutron stars via neutrino emission and structural change due to the softening of nuclear equation of state (EOS)\cite{kn86,t95,l96,pbpelk97}. 
Stimulated by the absence of a pulsar signature in SN 1987A, a delayed collapse scenario has been proposed in which a first-order phase transition of a protoneutron star to a kaon-condensed neutron star is assumed\cite{bb94}. According to this scenario, if the mass of a  protoneutron star exceeds the maximum mass of a kaon-condensed neutron star ($\sim 1.5 M_\odot$),  the star becomes gravitationally unstable after the deleptonization or the initial cooling stage due to a significant softening of the EOS, and finally collapses into a low-mass black hole. A numerical simulation of a delayed collapse from a hot neutron star to a black hole was performed by Baumgarte et al.\cite{bst96}, where the nuclear EOS with kaon condensation at zero temperature was used. \footnote{
Other delayed collapse scenarios related to a transition to pion condensation\cite{tt88,ts88}, hyperon-mixed matter\cite{g95,kj95,p99}, or quark matter\cite{bh89,pcl95} have also been considered by several authors. } 

Recently,  a delayed collapse of protoneutron stars has been studied by the use of the EOS of the kaon condensed phase at finite temperature which includes thermal effects and neutrino degeneracy\cite{ty98,yt01,p00,p01}. 
In such an early stage of neutron star evolution, where temperature is high (several tens of MeV), the mean free paths of neutrinos are less than the radius of a neutron star. Then highly dense hadronic matter at high temperature of several tens of MeV is opaque to neutrinos, and neutrinos are trapped and become degenerate in stars. 
Yasuhira and Tatsumi have shown that not only thermal effects on the free energy but also neutrino degeneracy delays an onset of kaon condensation, and that the latter further suppresses the subsequent development of condensates, which causes the large modification of the EOS before and after the  deleptonization stage\cite{yt01}. Based on the static properties of protoneutron stars, they discussed gravitational stability of the protoneutron stars accompanying a phase transition to a kaon-condensed star during the deleptonization era. 
 They obtained the range of the gravitational mass in which a delayed collapse occurs. 

There are some typical time scales associated with a delayed collapse of protoneutron stars. 
It has been shown in Ref.~\cite{mti00} that a time scale for an appearance of kaon condensation and its subsequent growth through nonequilibrium weak process, $nn\rightarrow npK^-$, is negligible as compared with the  time scales for deleptonization and initial cooling of order of ten seconds\cite{mti00}. A time delay of a collapse of protoneutron stars is mainly caused by deleptonization and initial cooling. 
In order to know the mechanisms of a delayed collapse in detail and to estimate the deleptonization and cooling time scales, we have to pursue dynamical and thermal evolution of a protoneutron star by taking into account neutrino transport caused by neutrino diffusion. The neutrino opacity (the inverse of the mean free path) plays an important role in  neutrino transport\cite{pls01}: 
During the deleptonization era, neutrinos are highly degenerate with their chemical potential being of order of 200 $-$ 300 MeV, and the total electron-lepton number fraction $Y_{le}$ (=$Y_e+Y_{\nu_e}$) amounts to 0.3$-$0.4. In this case, deleptonization proceeds through the neutrino chemical potential diffusion such that 
\beq
\frac{\partial Y_{le}}{\partial t}\propto\frac{1}{r^2}\frac{\partial}{\partial r}\Bigg\lbrack r^2 (D_2+D_{\bar 2}) \frac{\partial(\mu_{\nu_e}/T)}{\partial r}\Bigg\rbrack \ , 
\label{eq:cdiffusion}
\eeq
where $D_2$ and $D_{\bar 2}$ are typical diffusion constants which are related to the Rosseland mean of the neutrino opacities\cite{rpl98}.\footnote{Throughout this paper, the units $\hbar=c=k_B=1 $ are used.}
 Subsequently, 
during the initial cooling era, neutrinos become nondegenerate  ($\mu_{\nu_e}\simeq 0$), and matter cools down through neutrino diffusion driven by the temperature gradient such that 
\beq
C_{\rm V}\frac{\partial T}{\partial t}=\frac{7}{12}\sigma_{\rm SB} \frac{1}{r^2} \frac{\partial}{\partial r} \Bigg(r^2 T^3  \lambda^R \frac{\partial T}{\partial r}\Bigg) \ , 
\label{eq:dtdt}
\eeq
where $C_{\rm V}$ is the specific heat , $\sigma_{\rm SB}$ the Stefan-Boltzmann constant, and $\lambda^R$ is the Rosseland mean free path\cite{rpl98,ss79}. 

The neutrino opacity originates from absorption and scattering processes. In a normal phase, 
the neutrino absorption process $\nu_en\rightarrow e^- p $ as a one-nucleon process (we call this process the direct absorption process throughout this paper, and abbreviate it to DA) is forbidden in case the temperature is less than a few tens of MeV. This is because the phase space is exponentially small for those degenerate nucleons that satisfy energy-momentum conservation.\footnote{In such a rather low temperature case, the kinematical condition for occurrence of DA is the same as that for the direct Urca process, $n\rightarrow p e^- \bar\nu_e$, $p e^-\rightarrow n \nu_e$, which is relevant to a neutron star cooling of ordinary evolutionary stage. This kinematical condition depends on the baryon density dependence of the nuclear symmetry energy which remains controversial\cite{lpph91,fmtt94}. } Therefore, 
a spectator nucleon ($N$) participates in order to satisfy the kinematical condition, and the neutrino is absorbed through the two-nucleon process $\nu_e nN\rightarrow e^- p N $ ($N=p,n$)\cite{ss79,fm79,hj87}. In case the incident neutrino energy $E_\nu$ is large or temperature is high enough, DA becomes possible. Some authors considered DA for the neutrino absorption processes\cite{rpl98,rplp99} and found that DA has a contribution to the neutrino mean free path comparable to the neutrino-nucleon scattering (S) process, $\nu_e N\rightarrow \nu_e N$, which has  been considered as a main weak reaction giving the neutrino opacity\cite{ss79,hj87,ip82}. 
 
In the presence of kaon condensates, the following neutrino absorption process becomes possible as a {\it unique} process under the background of kaon condensates: 
\beq
\nu_e(\bfpnu)+N(\bfpi)\rightarrow e^- (\bfpe)+N(\bfpf) \ \ (N=p, n) \ , 
\label{eq:ab}
\eeq
which may be written symbolically as 
$\nu_e N \langle K^-\rangle\rightarrow e^- N$ 
with $ \langle K^-\rangle$ being the classical kaon field. \footnote{If the $p$-wave kaon-baryon interactions are take into account, it is possible that $p$-wave kaon condensation may be realized from neutron-star matter accompanying hyperon excitations\cite{m93,kvk95} or from hyperonic matter\cite{m02}. Throughout this paper, we concentrate on the primary mechanisms of neutrino absorption process in kaon condensates, so that we simply consider the $s$-wave kaon condensation realized from protoneutron-star matter. } 
Throughout this paper, we call this process the kaon-induced absorption process, and abbreviate it to KA. In Ref.~\cite{p01}, the neutrino opacity for KA has been briefly mentioned. 

In this paper, we consider KA in detail and obtain the neutrino mean free path $\lambda$(KA) for KA in both the nondegenerate ($\mu_\nu$=0) and degenerate ($\mu_\nu>0$) neutrino cases. 
We use the EOS with kaon condensation at finite temperature\cite{ty98,yt01} within a framework of chiral symmetry, which makes possible a consistent calculation of the EOS and neutrino processes. We compare $\lambda$(KA) with other relevant mean free paths such as DA and S, and discuss the importance of KA in neutrino transport phenomena during the dynamical and thermal evolution of protoneutron stars including a phase transition to a kaon-condensed phase.  

In Sec.~\ref{sec:chiral}, we overview the chiral symmetry approach to kaon condensation. In Sec.~\ref{sec:nondege}, the mean free paths for KA and S are derived in the nondegenerate neutrino case, and the numerical results are presented. The mean free paths are obtained and compared numerically in the degenerate neutrino case in Sec.~\ref{sec:dege}. In Sec.~\ref{sec:cooling}, rough estimates of the cooling time scale due to neutrino diffusion are given including the KA and S processes for a neutron star with kaon condensates. Summary and concluding remarks are given in Sec.~\ref{sec:summary}. 

\section{Chiral symmetry approach to kaon condensation }
\label{sec:chiral}

We first give an outline of the kaon-condensed state on the basis of 
chiral symmetry. In the framework of $SU_{\rm L}(3)\times SU_{\rm R}(3)$ current 
algebra and PCAC, the $s$-wave $K^-$ condensed state,  
$|K^-\rangle$, is generated by a chiral rotation of 
the normal state $|0\rangle$ 
as $|K^-\rangle=\hat U_K |0\rangle$,  
with the unitary operator $\hat U_K$ given by 
\begin{equation} 
\hat U_K\equiv\exp(i\mu_K t Q_{em}) \exp(i\theta Q_4^5) \ , 
\label{uk}
\end{equation} 
where $\mu_K$ is the kaon chemical potential, $\theta$ the chiral 
angle which represents the order parameter of the system, and
$Q_{em}$ ($Q_4^5$) the electromagnetic charge (the axial-vector charge).  
The classical $K^-$ field is  then 
given as \begin{equation} 
\langle \hat K^- \rangle\equiv \langle K^- |\hat K^-|K^-\rangle
=\langle 0|\hat U_K^{-1}\hat K^- \hat U_K|0\rangle
=\frac{f}{\sqrt 2}\sin\theta \exp(+i\mu_K t) \ , 
\label{kfield} 
\end{equation} 
where $f$ is the meson decay constant. 
Here we take the numerical value of $f$ to be the pion decay
constant (=93 MeV), following the previous papers instead of the kaon decay constant(=113 MeV),
which amounts to taking the lowest-order value 
in the chiral perturbation theory. 

Recently this framework has been extended to take into account the
quantum and/or thermal fluctuations around the condensate,
in accordance with chiral symmetry \cite{ty98}. 
The effective partition function
$Z_{chiral}$ at temperature $T$ can be formally written in the form of the 
imaginary-time path integral ($\beta=1/T$):
\beq
  Z_{chiral}=N\int [dU][dB][d\bar B] \exp\left[\int_0^\beta d\tau\int d^3x
   \left\{{\cal L}_{chiral}(U, B)+\delta{\cal L}(U, B)\right\}\right],
  \label{eq:part}
\eeq
where ${\cal L}_{chiral}(U, B)={\cal L}_0(U, B)+{\cal
L}_{SB}(U, B)$ being the Kaplan-Nelson Lagrangian\cite{kn86}
and is represented with 
the octet baryon field, $B$, and the chiral field, 
$U=\exp[i\lambda_a\phi_a/f]\in SU(3)$,  with 
the Goldstone fields $\phi_a$. $\delta{\cal L}$ is the induced SB term as
a result of introduction of chemical potentials,
$\mu_K (=\mu_Q: {\rm the ~charge ~chemical ~potential})$ and the baryon chemical potential $\mu_B$,
\begin{eqnarray}
 \delta {\cal L}
  &=& 
  -\frac{f^2\mu_K}{4}{\rm tr}\{[T_{em}, U]
  \frac{\partial U^\dagger}{\partial\tau}+
  \frac{\partial U}{\partial\tau}[T_{em}, U^\dagger]\}
\nonumber\\
 &&{} -\frac{\mu_K}{2}{\rm tr}
  \{B^\dagger[(\xi^\dagger[T_{em}, \xi]+\xi[T_{em}, \xi^\dagger]), B]\}
\nonumber\\
 &&{}-\frac{f^2\mu_K^2}{4}{\rm tr}\{[T_{em}, U][T_{em}, U^\dagger]\}
  +\mu_B{\rm tr}\{B^\dagger B\}-\mu_K{\rm tr}\{B^\dagger[T_{em},B]\} 
\label{del}
\end{eqnarray}
with $U=\xi^2$
and $T_{em}={\rm diag}(2/3, -1/3, -1/3)$.

The
local coordinates around the condensed point on the chiral manifold 
are introduced by the
following parameterization for $U$;
\beq
  U=\zeta U_f\zeta(\xi=\zeta U_f^{1/2} u^\dagger=u U_f^{1/2}\zeta),
   \quad \zeta=\exp(i\langle M\rangle/\sqrt{2}f),
  \label{eq:para}
\eeq
where $\langle M\rangle$ represents the condensate, 
$\langle M\rangle=V_+\langle K^+\rangle+V_-\langle K^-\rangle$, with 
$K^{\pm}=(\phi_4\mp i\phi_5)/\sqrt{2}$ and 
the V-spin operators $V_\pm=(\lambda_4\pm i\lambda_5)/2 $, while 
$U_f =\exp[i\lambda_a\phi_a/f]$ means the fluctuation field. The chiral angle $\theta$ is related to the classical fields $\langle K^\pm \rangle$ through the equation  
$\theta^2= 2\langle K^+\rangle \langle K^-\rangle/f^2$. Here we can easily see that the chiral 
transformation in $\hat U_K$, $\exp(i\theta Q_4^5)$, 
exactly corresponds to $\zeta$ in this framework.
Accordingly,
defining a new baryon field $B'$ by way of
\beq
  B'=u^\dagger B u,
\eeq
we can see that 
\begin{eqnarray}
  {\cal L}_{chiral}(U, B)
	&=&
  {\cal L}_0(U_f, B')+{\cal L}_{SB}(\zeta U_f\zeta,u B'
  u^\dagger),\nonumber \\
  \delta{\cal L}(U, B)
	&=&\delta{\cal L}(\zeta U_f\zeta,u B'
  u^\dagger).
\label{sbterm}
\end{eqnarray}
Thus only the SB terms {\it prescribe} the $KN$ dynamics in the condensed
phase; we can easily see that
the $KN$ sigma terms, $\Sigma_{Ki} (i=n,p)$, stem from ${\cal
L}_{SB}$, while the Tomozawa-Weinberg term from $\delta{\cal L}$
\cite{ty98}.

We then evaluate the partition function $Z_{chiral}$.
During this course, there appear the excitation spectra of kaonic modes
with energies $(E_\pm^K)$.
The mode with $E_-^K$ is the Goldstone mode and exhibits the Bogoliubov
spectrum,
which stems from the spontaneous breaking of $V$-spin
symmetry in the condensed phase \cite{ty98}. 
Under the relevant approximation,
they are reduced to the simple form
\beq
  E_\pm^K({\bf p})
  \simeq
	\sqrt{p^2+m_K^{*2}\cos\theta +b^2}
	\pm(b+\mu_K\cos\theta),
 \label{eq:dispa}
\eeq
where $m_K^{\ast 2} = m_K^2 - \rho_B/f^2 (\Sigma_{Kp}x+\Sigma_{Kn}(1-x))$ with $m_K^\ast$
being the effective mass of kaons and proton fraction $x=\rho_p/\rho_B$ , 
and $b=\rho_B(1+x)/(4f^2)$, which stems from the Tomozawa-Weinberg term representing the 
$KN$ $s$-wave interaction.

Eventually the effective thermodynamic potential 
$\Omega_{chiral}=-T\ln Z_{chiral}$ reads, up to the one-loop order,  
\beq
  \Omega_{chiral}=\Omega_c+\Omega_K^{th}+\Omega_N,
   \label{eq:omega}
\eeq
where $\Omega_c$ is the classical kaon contribution,
\beq
  \Omega_c=V[-f^2m_K^2(\cos\theta-1)
   -1/2\cdot\mu_K^2f^2\sin^2\theta],
  \label{eq:omegac}
\eeq
and the thermal kaon contribution 
is given as follows;
\beq
  \Omega_K^{th}=TV\int \frac{d^3 p}{(2\pi)^3}
  \ln(1-e^{-\beta E_+^K({\bf p})})(1-e^{-\beta E_-^K({\bf p})}).
\eeq
It is to be noted that the zero-point-energy contribution of kaons is
very tiny \cite{ty98}\cite{te97}
and we may discard it.
$\Omega_N$ denotes the nucleon contribution,
\beq
  \Omega_N
    = \Omega^{kin}_N + \Omega^{pot}_N, 
\eeq
which consists of the kinetic and potential contributions; 
\begin{eqnarray}
  \Omega_N^{kin}
    &=& {}-2TV\sum_{n,p}\int\frac{d^3p}{(2\pi)^3}
	\ln(1+e^{-\beta(E_i({\bf p})-\mu_i)}),
	\nonumber\\
  \Omega^{pot}_N
    &=& {}-\rho_B\frac{\partial E^{pot}_N}{\partial \rho_B}
	\label{eq:npot}
\end{eqnarray}
with the potential energy $E^{pot}_N$. $\mu_i (i=p, n)$ are the chemical 
potentials of nucleons: $\mu_n=\mu_B, \mu_p=\mu_B-\mu_K$ under chemical 
equilibrium.
The single-particle energies for nucleons are given as  
\begin{eqnarray}
  E_p({\bf p})
    &=& \frac{p^2}{2m_N} - \left(\Sigma_{Kp}+\mu_K\right)
	\left( 1-\cos\theta\right)
	+ \frac{1}{V}\frac{\partial E^{pot}_N}{\partial \rho_p},
	\nonumber\\
  E_n({\bf p})
    &=& \frac{p^2}{2m_N} - \left(\Sigma_{Kn}+\frac{\mu_K}{2}\right)
	\left(1-\cos\theta\right)
	+ \frac{1}{V}\frac{\partial E^{pot}_N}{\partial \rho_n}.
  \label{eq:single}
\end{eqnarray}
Sometimes it is convenient to use the shifted chemical potentials, $\mu^0_i 
(i=p, n)$, defined by
\begin{eqnarray}
\mu^0_p&=&\mu_p+\left(\Sigma_{Kp}+\mu_K\right)
	\left( 1-\cos\theta\right)
-\frac{1}{V}\frac{\partial E^{pot}_N}{\partial \rho_p},
	\nonumber\\
\mu^0_n&=&\mu_n+\left(\Sigma_{Kn}+\mu_K/2\right)
	\left(1-\cos\theta\right)
	- \frac{1}{V}\frac{\partial E^{pot}_N}{\partial \rho_n},
\label{shift}
\end{eqnarray} 
which make the $\Omega_N^{kin}$ in Eq. (\ref{eq:npot}) to be in a simple form,
\beq
\Omega_N^{kin}
    = -2TV\sum_{n,p}\int\frac{d^3p}{(2\pi)^3}
	\ln(1+e^{-\beta(p^2/2m_N-\mu_i^0)})
\label{shifto}
\eeq 

Here, following Prakash et al.\cite{ains},  we have introduced 
the effective potential-energy for nucleons in the form,
\beq
  E^{pot}_N = E^{sym}_N + E^{V}_N,
\eeq
both of which cannot be given by the Kaplan-Nelson Lagrangian.
$E^{sym}_N$ represents the symmetry energy contribution,
\beq
  E^{sym}_N
    = V\rho_B(1-2x)^2 S^{pot}(u); 
\quad
(u = \rho_B/\rho_0; \rho_0 =
0.16{\rm fm}^{-3})
  \label{eq:sym}
\eeq
where the function $S^{pot}(u)$ reads,
\beq
  S^{pot}(u)=(S_0-(2^{2/3}-1)(3/5)E_F^0)F(u)
\eeq
with the constraint $F(u=1)=1$
to reproduce the empirical symmetry energy $S_0\simeq 30$MeV 
at the nuclear density $\rho_0$.
$E_F^0$ is the Fermi energy at $\rho_0$.
Hereafter we use $F(u)=u$ for an example. 
The residual potential contribution $E^{V}_N$ renders
\beq
  E^{V}_N/V
    = \frac{1}{2} A u^2 \rho_B 
	+ \frac{B u^{\sigma+1}\rho_B}{1+B'u^{\sigma-1}}
 	+ 3 u \rho_B \sum_{i=1,2} C_i \left(
	\frac{\Lambda_i}{p_F^0} \right)^3
	\left( \frac{p_F}{\Lambda_i}
	- \arctan \frac{p_F}{\Lambda_i} \right).
  \label{eq:vu}	
\eeq
We choose the parameter set
($A$,$B$,$B'$,$\sigma$,$C_i$,$\Lambda_i$)
for compression modulus to be $K_0=240$MeV for normal nuclear matter.

Using the thermodynamic relations,
\beq
 S_{chiral}=-\frac{\partial\Omega_{chiral}}{\partial T}, \quad
  Q_i=-\frac{\partial\Omega_{chiral}}{\partial\mu_i}, \quad
  E_{chiral}^{int}=\Omega_{chiral}+TS_{chiral}+\sum\mu_iQ_i,
 \label{eq:rel}
\eeq
we  can find entropy, charge and internal energy, respectively. 

The total thermodynamic potential  $\Omega_{total}$ 
is given by adding the one for
leptons (electrons, muons and neutrinos),
$\Omega_l$, $\Omega_{total}=\Omega_{chiral}+\Omega_l$; 
\beq
\Omega_l=-2TV\sum_{e,\mu, \nu}\int\frac{d^3
p}{(2\pi)^3}\left[\ln(1+e^{-\beta(E_i({\bf p})-\mu_i)})
+\ln(1+e^{-\beta(E_{\bar i}({\bf p})+\mu_i)})\right],
\label{omegal}
\eeq
with $E_i({\bf p})=E_{\bar i}({\bf
p})=\sqrt{m_i^2+p^2}$
($i=e,\mu,\nu_e,\nu_\mu$). 

\section{Nondegenerate neutrino case}
\label{sec:nondege}

In the nondegenerate case, the neutrino chemical potential may be set equal to zero, so that the abundance of neutrinos and antineutrinos are small, ($Y_\nu$, $Y_{\bar\nu}\ll 1)$. The antineutrino processes, $\bar\nu_e N e^-\rightarrow N\langle K^-\rangle $,  $ \bar\nu_e N\rightarrow \bar\nu_e N$, have an equal contribution to the reaction rates with the neutrino processes, so that we here consider only the neutrino processes. 

\subsection{Kaon-induced neutrino absorption process (KA) }
\label{subsec:absorption}

The inverse of the mean free path,  for (\ref{eq:ab}), consists of two 
contributions: 
\begin{equation}
1/\lambda({\rm KA})=\sum_{N=p, n}1/\lambda^{(N)}({\rm KA}), 
\label{opactotal}
\end{equation}
where $\lambda^{(N)}({\rm KA})$ is the mean-free path for the process, 
$\nu_e N\langle K^-\rangle\rightarrow e^- N$. 
Each mean free path is given as 
\beqa
1/\lambda^{(N)}({\rm KA})&=&\frac{1}{V}\frac{V^3}{(2\pi)^9}\int d^3 p_N \int d^3 p_e \int d^3 p_{N}' \delta(E_e+E_{N}'-E_\nu-E_N-\mu_K) \cr
& &\times S_a W_{\rm fi,a}^{(N)} \ \ \ ({\rm for} \ N=p,n) \ , 
\label{eq:lka}
\eeqa
where $V$ is the normalization volume,  $S_a\equiv f(\bfpi)[1-f(\bfpe)][1-f(\bfpf)]$ being the statistical factor with the Fermi-Dirac distribution function $f({\bf p}_i)=1/[\exp\{(E_i-\mu_i)/T\}+1]  $, and $W_{\rm f i,a}^{(N)}$ is the transition rate: 
\beq
W_{\rm f i, a}^{(N)}=(2\pi)^3 V\delta^{(3)}(\bfpe+\bfpf-\bfpnu-\bfpi)\sum_{\text{spins}}|M_{\rm a}^{N}|^2 \cdot\frac{1}{V^3} \ , 
\label{tm}
\eeq
with the squared matrix element $ \displaystyle\sum_{\text{spins}}|M_{\rm a}^{N}|^2$. 
The summation is taken over the initial and final nucleon spins. 
One can see from the energy delta function in Eq.~(\ref{eq:lka}) that the system is supplied with the energy through the kaon chemical potential $\mu_K$, which comes from the time dependence of the classical kaon field (\ref{kfield}). KA has a close connection to the kaon-induced Urca (KU) process, 
\begin{mathletters}\label{ku}
\beqa
& & N\langle K^- \rangle\rightarrow Ne^-  \bar\nu_e \ , 
 \label{eq:kuf} \\
& & Ne^-\rightarrow N\langle K^-\rangle\nu_e \ ,  
 \label{eq:kub}
\eeqa
\end{mathletters}
which gives a rapid cooling mechanism of kaon-condensed neutron stars\cite{b88,t88}: 
The mean free path for KA for the nondegenerate neutrino case is related to the neutrino emissivity $E^{(N)}({\rm KU\textrm{-}F})$ for the forward process of KU [ (\ref{eq:kuf}), denoted as KU-F ] as
\beq
E^{(N)}({\rm KU\textrm{-}F})=\frac{1}{(2\pi)^3}\int d^3 p_\nu E_\nu\cdot 1/\lambda^{(N)}({\rm KA})(-E_\nu, T) \ . 
\label{eq:rel}
\eeq

\subsubsection{Squared matrix elements}
\label{subsubsec:sqmatrix}

 The effective weak Hamiltonian is of the current$-$ 
current interaction type,  \\ 
$\displaystyle H_{\rm W}={G_F\over \sqrt 2}
 J_{\rm h}^\mu\cdot l_\mu+\text{h.c.}$ with $l_\mu$ [=$\bar\psi_e \gamma_\mu (1-\gamma_5)\psi_\nu $  ] being the charged leptonic current and the charged hadronic current,  
$J_{\rm h}^\mu=\cos\theta_{\rm C} (V_{1+i2}^\mu -A_{1+i2}^\mu)
+\sin\theta_{\rm C} (V_{4+i5}^\mu -A_{4+i5}^\mu) $, 
where $\theta_{\rm C}(\simeq 0.24)$ is the Cabibbo angle,  
and $V_a^\mu$ and $A_a^\mu$ are the vector and axial-vector currents, 
respectively. 
In the kaon-condensed state $|K^-\rangle$, 
the matrix elements are given 
 from the transformed Hamiltonian\cite{t88,fmtt94}  
\begin{equation}
 {\widetilde H}_{\rm W}=\hat U_K^{-1}H_{\rm W}\hat U_K 
={G_{F}\over \sqrt 2}
{\widetilde J}_{\rm h}^\mu l_\mu
+\text{ h.c.} \ ,
\label{th}
\end{equation}
where the effective hadronic current ${\widetilde J}_{\rm h}^\mu$ 
is given in a model-independent way by the use of current algebra: 
\begin{eqnarray}
 {\widetilde J}^\mu_{\rm h}
&=&\hat U_K^{-1}J^\mu_{\rm h}\hat U_K \cr
&=&e^{-i\mu_K t}\Bigg\lbrack\cos\theta_c\bigg\{
(V_{1+i2}^\mu-A_{1+i2}^\mu)\cos(\theta/2)
+i(V_{6-i7}^\mu-A_{6-i7}^\mu)\sin(\theta/2)\bigg\} \cr
&+&\sin\theta_c\bigg\{(V_4^\mu-A_4^\mu)+i\cos\theta(V_5^\mu-A_5^\mu)
-i\sin\theta\Big(V_V^\mu-A_V^\mu\Big)\bigg\}\Bigg\rbrack \ ,  
\label{tjh}
\end{eqnarray} 
where $\displaystyle V_V^\mu\equiv\frac{1}{2}(V_3^\mu+\sqrt{3} V_8^\mu) $  and 
$\displaystyle A_V^\mu\equiv\frac{1}{2}(A_3^\mu+\sqrt{3} A_8^\mu) $. 

The spin-summed squared matrix element is then written as 
\beq
\sum_{\text{spins}}|M_{\rm a}^{N}|^2 =\frac{G_F^2}{2}H_{\mu\nu}^{(N)}({\rm KA}) L^{\mu\nu} \, 
\label{eq:ma}
\eeq
where $H_{\mu\nu}^{(N)}({\rm KA})$ is the hadronic tensor for KA : 
\beq
H_{\mu\nu}^{(N)}({\rm KA})=\sum_{s,s'=\pm 1/2}\langle N',s'|{\widetilde J}_{{\rm h},\mu} | N,s\rangle \langle N,s|{\widetilde J}_{{\rm h},\nu}^\dagger | N',s' \rangle \ , 
\label{eq:hmna}
\eeq
and $L^{\mu\nu}$ the leptonic tensor:
\beq
L^{\mu\nu}=2\Big\lbrack (p_\nu)^\nu (p_e)^\mu+
(p_\nu)^\mu (p_e)^\nu-(p_\nu\cdot p_e) g^{\nu\mu}-i\epsilon^{\mu\nu\eta\zeta}(p_\nu)_\eta (p_e)_\zeta \Big\rbrack/(p_\nu)^0 (p_e)^0 \ . 
\label{eq:leptona}
\eeq
For KA, the matrix elements in the hadronic tensor (\ref{eq:hmna}) come from the last term in the transformed hadronic current (\ref{tjh}), which is proportional to $\sin\theta$ . 

For nonrelativistic nucleons, the vector  and axial-vector currents are reduced to 
\begin{eqnarray}
V_3^\mu&=&\frac{1}{2}\tau_3\delta^\mu_0 \ , A_3^\mu=\frac{g_A}{2}\tau_3\sigma^i\delta^\mu_i \cr
V_8^\mu&=&\frac{\sqrt{3}}{2}Y\delta^\mu_0 \ , A_8^\mu=\frac{\sqrt{3}}{2}Y{\tilde g}_A\sigma^i\delta^\mu_i \qquad(\textrm{for}\quad i=1-3)
\label{nonrela}
\end{eqnarray}
with $Y$ being the hypercharge ($Y$=1 for the nucleon), 
$g_A(=D+F=1.25)$, and 
${\tilde g}_A(=F-D/3 =0.17) $ the axial-vector coupling strengths, 
where $D$=0.81, $F$=0.44. It is to be noted that the matrix elements 
$H_{\mu\nu}^{(N)}({\rm KA})$ are 
the same as those for KU. We refer to Refs.~\cite{fmtt94,t88} for the expression of $H_{\mu\nu}^{(N)}({\rm KU})$. 

After taking the angular average of the reduced squared matrix element $H_{\mu\nu}^{(N)}({\rm KA})L^{\mu\nu} $ and assuming that its momentum dependence is small, one obtains 
\begin{mathletters}\label{eq:matrixa}
\beqa
\overline{H_{\mu\nu}^{(p)}({\rm KA}) L^{\mu\nu}}&=&
\frac{1}{4}\sin^2\theta_{\rm C}\sin^2
\theta\Big\lbrace 16+3(g_A+3\widetilde{g}_A)^2\Big\rbrace \ , 
\label{eq:matrixap} \\
\overline{H_{\mu\nu}^{(n)}({\rm KA}) L^{\mu\nu}}&=&
\frac{1}{4}\sin^2\theta_{\rm C}\sin^2
\theta\Big\lbrace 4+3(g_A-3\widetilde{g}_A)^2\Big\rbrace \ . 
\label{eq:matrixan}
\eeqa
\end{mathletters}
The expression for the mean free path is reduced to 
\beq
1/\lambda^{(N)}({\rm KA})=2\pi \frac{V^3}{(2\pi)^9}\frac{G_F^2}{2}\frac{(2\pi)^3}{V^3}\overline{H_{\mu\nu}^{(N)}({\rm KA}) L^{\mu\nu}} I^{(N)}({\rm KA})\ , 
\label{eq:la}
\eeq
where $I^{(N)}$(KA) is the phase space integral, the expression of which is given in Appendix~\ref{sec:phase}.  
The phase space integral $I^{(N)}$(KA) may be performed numerically. 

\subsubsection{Low-temperature approximation }
 \label{subsubsec:lowt}
 
 The phase space integral $I^{(N)}$(KA) [see Eq.~(\ref{eq:i})] is rewritten in the form
 \beqa
 I^{(N)}({\rm KA})=\widetilde{m_N}^2\frac{T^5}{\widetilde{E}_\nu }
 & &\int_{-\widetilde\mu_N}^\infty d x_2\int _{-\infty}^{\widetilde\mu_N} d x_4(x_2+x_4+\widetilde{E}_\nu + \widetilde\mu_K) \cr
 &\times& \frac{1}{[\exp(x_2)+1]}\frac{1}{[\exp(x_4)+1]}
 \frac{1}{[\exp\{-(x_2+x_4+ \widetilde{E}_\nu)\}+1]}F_{\rm a}\ , 
 \label{eq:i2}
 \eeqa
 where ` $\sim$ ' stands for a dimensionless quantity divided by the temperature $T$ (e.g., $ \widetilde{E}_\nu\equiv E_\nu/T$ ), 
 and $x_2\equiv \widetilde{E}_N-\widetilde\mu_N$, $ x_4\equiv -\widetilde{E}_{N} '+\widetilde\mu_N$ with $\mu_N$ being the chemical potential of the nucleon $N$. 
 
For low-temperature cases, analytic formula for the mean-free path $\lambda^{(N)}({\rm KA})$ 
is obtained after the following approximations are made: 
(i) $ \widetilde\mu_N\rightarrow \infty$ in the lower and upper bounds in the integrals in Eq.~(\ref{eq:i2}) since $T\ll \mu_N$. (ii) $\magpnu=E_\nu\ll \magpi,\magpf,\magpe$ since the neutrino is thermal [$E_\nu=O(T)$]. In this limit, $F_{\rm a}\rightarrow 8\pi^2 \widetilde{E}_\nu $. (iii) The factor $ x_2+x_4+\widetilde{E}_\nu + \widetilde\mu_K$ in the integrand of Eq.~(\ref{eq:i2}),  which is equal to $\widetilde{E}_e$ due to energy conservation, is replaced by $\widetilde\mu_e$ (=$\mu_e/T$). 

The result is 
\begin{equation}
1/\lambda^{(N)}({\rm KA})(E_\nu, T)=\frac{G_F^2}{16\pi^3}
\overline{H_{\mu\nu}^{(N)}({\rm KA}) L^{\mu\nu}} \cdot
m_N^2 \mu_e T^2\frac{\pi^2+\widetilde{E}_\nu^2}{1+e^{-\widetilde{E}_\nu}} \ . 
\label{eq:opac-pn} 
\end{equation}
By the substitution of Eq.~(\ref{eq:matrixa}), $\lambda_p$ and $\lambda_n$ are estimated as 
\begin{mathletters}\label{eq:analytic}
\begin{eqnarray}
1/\lambda^{(p)}({\rm KA})&=&(6.14\times10^{-5})\sin^2\theta\frac{\mu_e}{m_\pi}\Bigg(\frac{T}{1 {\rm
MeV}}\Bigg)^2
\frac{\pi^2+\widetilde{E}_\nu^2}{1+e^{-\widetilde{E}_\nu}} \quad [ {\rm m}^{-1} ] \ , 
\label{eq:opacity-p} \\
1/\lambda^{(n)}({\rm KA})&=&(1.37\times10^{-5})\sin^2\theta\frac{\mu_e}{m_\pi}\Bigg(\frac{T}{1 {\rm
MeV}}\Bigg)^2
\frac{\pi^2+\widetilde{E}_\nu^2}{1+e^{-\widetilde{E}_\nu}} \quad [ {\rm m}^{-1} ] 
\label{eq:opacity-n} \ , 
\end{eqnarray}
\end{mathletters}
where the electron chemical potential $\mu_e$ is divided by the pion mass  $m_\pi$ (=140 MeV) since numerically $\mu_e=O(m_\pi) $. From Eqs.~(\ref{eq:la}), (\ref{eq:i}), and (\ref{eq:angular}), one can see that a factor $T^2$ in Eq.~(\ref{eq:analytic}) comes from the energy integral for the electron and incoming and outgoing nucleons ($T^3$) with the energy-conserving delta function ($T^{-1}$).  
In general, the temperature dependence of the phase-space integral [ Eq.~(\ref{eq:i2}) ] is complicated. 
However,  the analytic formulae Eqs.~(\ref{eq:opacity-p}) and (\ref{eq:opacity-n}) in the low-temperature approximation 
have a simple scaling property, 
\beq 
1/\lambda^{(N)}({\rm KA})\Big(\frac{T}{T_0}E_\nu,T\Big)=\Big(\frac{T}{T_0}\Big)^2 \cdot 1/\lambda^{(N)}({\rm KA})(E_\nu, T_0) \ , 
\label{eq:scaling}
\eeq
as far as the temperature dependences of the parameters $\theta$, $\mu_e$ are neglected. \footnote{As seen in Table \ref{tab:ndparam}, the temperature dependence of the parameters are weak for $T\lesssim$ 40 MeV. See also \cite{yt01}. }
The scaling property (\ref{eq:scaling}) should be compared with that for the mean free path $\lambda$ for the absorption process, $\nu_e nn\rightarrow e^- pn $, in the normal neutron-star matter: 
$\displaystyle 1/\lambda\Big(\frac{T}{T_0}E_\nu,T\Big)=\Big(\frac{T}{T_0}\Big)^4\cdot 1/\lambda(E_\nu, T_0) $\cite{hj87}. 

\subsection{Neutrino scattering process (S) in kaon condensates}
\label{subsec:scattering}
  
In order to discuss a role of kaon-induced absorption processes in comparison 
with the neutrino-nucleon scattering process, we derive for the neutrino mean 
free path $\lambda^{(N)}({\rm S})$ due to the neutrino-nucleon scattering 
process
\beq
\nu_e(p_\nu) +  N(p_N)\rightarrow \nu_e(p_\nu ') +  N(p_N ') 
\qquad ({\rm for } \ N=p,n ) 
\label{eq:scattering}
\eeq
 in a kaon-condensed phase. 

The mean free path is given by 
\beq
1/\lambda^{(N)}({\rm S})=2\pi\frac{1}{V}\frac{V^3}{(2\pi)^9}\int d^3 p_N \int d^3 p_\nu ' \int d^3 p_N ' 
\delta(E_\nu '+E_N '-E_\nu -E_N ) S_{\rm s}W_{\rm f i,s}^{(N)}
\label{eq:ls}
\eeq
where $S_{\rm s}\equiv f (\bfpi)\Big\lbrack 1
-f (\bfpf)\Big\rbrack\Big\lbrack 1-f(\bfpnuf) \Big\rbrack $ is the statistical factor, and $W_{\rm f i,s}^{(N)}$ is the transition rate: 
\beq
W_{\rm f i, s}^{(N)}=(2\pi)^3 V\delta^{(3)}(\bfpnuf+\bfpf-\bfpnu-\bfpi)\sum_{\text{spins}}|M_{\rm s}^{N}|^2 \cdot\frac{1}{V^3} \ , 
\label{tm}
\eeq
with the squared matrix element $ \displaystyle\sum_{\text{spins}}|M_{\rm s}^{N}|^2$. The chemical potential $\mu_\nu$  appearing in the distribution function of the neutrino $f(\bfpnuf)$ is set equal to zero for nondegenerate neutrinos. 

\subsubsection{Squared matrix elements}
\label{subsubsec:sqmatrixs}

The scattering processes are mediated by the effective weak Hamiltonian of a type $\displaystyle H_{\rm W,Z}={G_F\over \sqrt 2}
 J_Z^\mu\cdot l_{Z,\mu}+\text{h.c.}$ , where $l_{Z, \mu}^\mu$ [=$\bar\psi_\nu \gamma_\mu(1-\gamma_5)\psi_\nu$] is the neutral leptonic current, and $J_Z^\mu$ is the 
 neutral hadronic current: 
\beqa
J_Z^\mu&=&h_3^\mu-2\sin^2\theta_{\rm W} h_{\rm em}^\mu \\
&=&(V_3^\mu-A_3^\mu)-2\sin^2\theta_{\rm W}\Big(V_3^\mu
+\frac{1}{\sqrt{3}}V_8^\mu\Big) \ . 
\label{hz}
\eeqa
where $\theta_{\rm W}$ is the Weinberg angle ($\sin^2\theta_{\rm W}\simeq 0.23$), and $ h_{\rm em}^\mu$ is the electromagnetic current. 
The chirally transformed hadronic current is given by\footnote{
In Ref.~\cite{jm98}, the neutral current processes for neutrino emissivities in kaon condensates have also been considered.  However, the resulting matrix element for the neutron process, $n\rightarrow n\nu\bar\nu$, is different from ours. }
\beqa 
\widetilde J_Z^\mu&=&\hat U_K^{-1}J_Z^\mu \hat U_K \cr
&=& V_3^\mu-A_3^\mu+\frac{1}{2}(\cos\theta-1)(V_V^\mu-A_V^\mu)
+\frac{1}{2}\sin\theta(V_5^\mu-A_5^\mu) \cr
&-&2\sin^2\theta_{\rm W}\Big\{V_3^\mu+\frac{1}{\sqrt{3}}V_8^\mu+(\cos\theta-1)V_V^\mu-\sin\theta A_5^\mu \Big\} \ . 
\label{trans}
\eeqa

The spin-summed squared matrix element for the scattering process 
(\ref{eq:scattering}) is then written as
\begin{equation}
|M_{\rm s}^{(N)}|^2=\frac{G_F^2}{2} H_{\mu\nu}^{(N)}({\rm S}) L_Z^{\mu\nu}
 \ , 
\label{matrix}
\end{equation}
where $H_{\mu\nu}^{(N)}({\rm S})$ is the hadronic tensor for S : 
\beq
H_{\mu\nu}^{(N)}({\rm S})=\sum_{s,s'=\pm 1/2}\langle N',s'|{\widetilde J}_{Z,\mu} | N,s\rangle \langle N,s|{\widetilde J}_{Z,\nu}^\dagger | N',s' \rangle \ , 
\label{eq:hmns}
\eeq
 and $L_Z^{\mu\nu}$ is the leptonic tensor for the incoming and outgoing neutrino pair :
\beq
L_Z^{\mu\nu}=2\Big\lbrack (p_\nu)^\nu (p_\nu ')^\mu+
(p_\nu)^\mu (p_\nu ')^\nu-(p_\nu\cdot p_\nu ') g^{\nu\mu}-i\epsilon^{\mu\nu\eta\zeta}(p_\nu)_\eta (p_\nu ')_\zeta \Big\rbrack/[(p_\nu)^0 (p_\nu ')^0)] \ . 
\label{eq:leptons}
\eeq
By the use of Eq.(\ref{nonrela}), each component of the hadronic tensor is given by 
\beqa
H^{(p)}_{00}({\rm S})&=&\frac{1}{2}(1-4\sin^2\theta_{\rm W})^2\cos^2\theta \cr
H^{(p)}_{0i}({\rm S})&=&0 \cr
H^{(p)}_{ij}({\rm S})&=&\frac{1}{2}\Big\{g_A-\frac{1}{4}(1-\cos\theta)(g_A+3{\tilde g}_A)\Big\}^2\delta^{ij} \ \ \ (i,j=1\sim 3)  
\label{hsp}
\eeqa
for $N=p$, and 
\beqa
H^{(n)}_{00}({\rm S})&=&\frac{1}{2}\Big\{1+\frac{1}{2}(1-4\sin^2\theta_{\rm W})(1-\cos\theta)\Big\}^2 \cr
H^{(n)}_{0i}({\rm S})&=&0 \cr
H^{(n)}_{ij}({\rm S})&=&\frac{1}{2}\Big\{g_A-\frac{1}{4}(1-\cos\theta)(g_A-3{\tilde g}_A)\Big\}^2\delta^{ij} \ \ \ (i,j=1\sim 3) 
\label{hsn}
\eeqa
for $N=n$. 

The reduced squared matrix elements are given by 
\beqa
H_{\mu\nu}^{(p)}({\rm S})L_Z^{\mu\nu}&=&(1-4\sin^2\theta_{\rm W})^2\cos^2\theta(1+\cos\theta_{13}) \cr
&+&\Big\{g_A-\frac{1}{4}(1-\cos\theta)(g_A+3{\tilde g}_A)\Big\}^2(3-\cos\theta_{13}) 
\label{hlsp}
\eeqa
for $N=p$, and 
\beqa
H_{\mu\nu}^{(n)}({\rm S})L_Z^{\mu\nu}&=&\Big\{1+\frac{1}{2}(1-4\sin^2\theta_{\rm W})(1-\cos\theta)\Big\}^2(1+\cos\theta_{13}) \cr
&+&\Big\{g_A-\frac{1}{4}(1-\cos\theta)(g_A-3{\tilde g}_A)\Big\}^2(3-\cos\theta_{13}) 
\label{hlsn}
\eeqa
for $N=n$, 
where $\theta_{13}$ is the angle between the momenta ${\bf p}_1$ and ${\bf p}_3$. 

By taking the angular average ($\overline{\cos\theta_{13}}\rightarrow 0$), one obtains 
\begin{mathletters}\label{eq:matrixs}
\beqa
\overline{H_{\mu\nu}^{(p)}({\rm S})L_Z^{\mu\nu}}
&=&(1-4\sin^2\theta_{\rm W})^2\cos^2\theta+3\Big\{g_A-\frac{1}{4}(1-\cos\theta)(g_A+3{\tilde g}_A)\Big\}^2 \ , \label{eq:matrixsp} \\
\overline{H_{\mu\nu}^{(n)}({\rm S})L_Z^{\mu\nu}}
&=&\Big\{1+\frac{1}{2}(1-4\sin^2\theta_{\rm W})(1-\cos\theta)\Big\}^2+3\Big\{g_A-\frac{1}{4}(1-\cos\theta)(g_A-3{\tilde g}_A)\Big\}^2 \ . 
\label{eq:matrixsn}
\eeqa
\end{mathletters}
In the limit $\theta\rightarrow 0$, the reduced squared matrix elements Eqs.~(\ref{eq:matrixsp}) and (\ref{eq:matrixsn}) tend to $(1-4\sin^2\theta_{\rm W})^2+3g_A^2$ and $1+3g_A^2$, respectively. Thus the scattering process is not unique to kaon condensation but is  operative even in the normal phase so far as the kinematical condition is  satisfied. 

\subsubsection{Phase-space integrals}
\label{subsubsec:phase}

After performing the phase-space integrals in (\ref{eq:ls}), one obtains 
\begin{equation}
1/\lambda^{(N)}({\rm S})=\frac{G_F^2}{2(2\pi)^5}
\overline{H_{\mu\nu}^{(N)}({\rm S}) L_{\rm Z}^{\mu\nu}}I^{(N)} ({\rm S})
\label{eq:finals}
\end{equation}
with 
\begin{equation}
I^{(N)}({\rm S})=\frac{1}{E_\nu}\int_0^\infty \magpi d\magpi
\int_0^\infty \magpf d\magpf E'_{\nu}\Theta(E'_{\nu})
f_N(\bfpi)[ 1-f_N(\bfpf) ]
F_{\rm s}(E_\nu,\magpi,E'_{\nu},\magpf) \ , 
\label{isN}
\end{equation}
where $E_\nu$ (=$\magpnu$) is the neutrino energy, $\Theta(x)=1$ for $x>0$, $\Theta(x)=0$ for $x<0$,  and $E_{\nu}'$ is related by 
\begin{equation}
E_{\nu}' \equiv \frac{\bfpi^2}{2m_N}-\frac{\bfpf^2}{2m_N}+E_\nu
\label{e3}
\end{equation}
The function $F_{\rm s}$ is the same as $F_{\rm a}$ in Eq.~(\ref{eq:angular}) but that $\magpe$ in the argument of $F_{\rm a}$ is to be replaced with $\magpnuf$ (=$E'_\nu$). 

\subsection{Numerical results in the nondegenerate neutrino case}
\label{subsec:ka}

We discuss the numerical results of the mean free path $\lambda({\rm KA})$ for  KA in the nondegenerate neutrino case. First we check the validity of the analytic formulae for $\lambda({\rm KA})$ obtained in the low-temperature approximation. Secondly we compare $\lambda({\rm KA})$  with the other dominant mean free paths for scattering processes in kaon condensates and those in the normal (noncondensed) neutron-star matter. The parameters needed for obtaining the results are listed in Table \ref{tab:ndparam}. 
These values have been obtained by Yasuhira and Tatsumi based on the EOS at finite temperature for kaon condensation in a neutrino-free case ($Y_{le}=0$) \cite{yt01}. Note that $\mu_p^0$ and $\mu_n^0$ in the table are not the naive chemical potentials for the nucleons, but they  appear in the Fermi distribution 
functions (see Eqs. (\ref{shift}), (\ref{shifto})). 
In this paper, the possible range of temperature $T$ of the kaon-condensed matter is taken to be 0$-$60 MeV. In the framework of Ref.~\cite{yt01}, the critical density for the onset of kaon condensation is $\rho_{\rm crit}$=0.49 fm$^{-3}$ at zero temperature and $\rho_{\rm crit}$=0.54 fm$^{-3}$ at $T$=60 MeV for a neutrino-free case. Thus the density $\rhob$=0.57 fm$^{-3}$ in Table~\ref{tab:ndparam} 
corresponds to the weaker condensation especially for high temperatures, while the density $\rhob$=0.72 fm$^{-3}$ corresponds to the stronger condensation over the relevant temperatures. 

\subsubsection{Comparison of numerically integrated results with  analytic formulae}
\label{subsubsec:1}

\begin{figure}[tt]\noindent
\begin{minipage}[l]{0.5\textwidth}
\centerline{
\epsfxsize=\textwidth
\epsffile{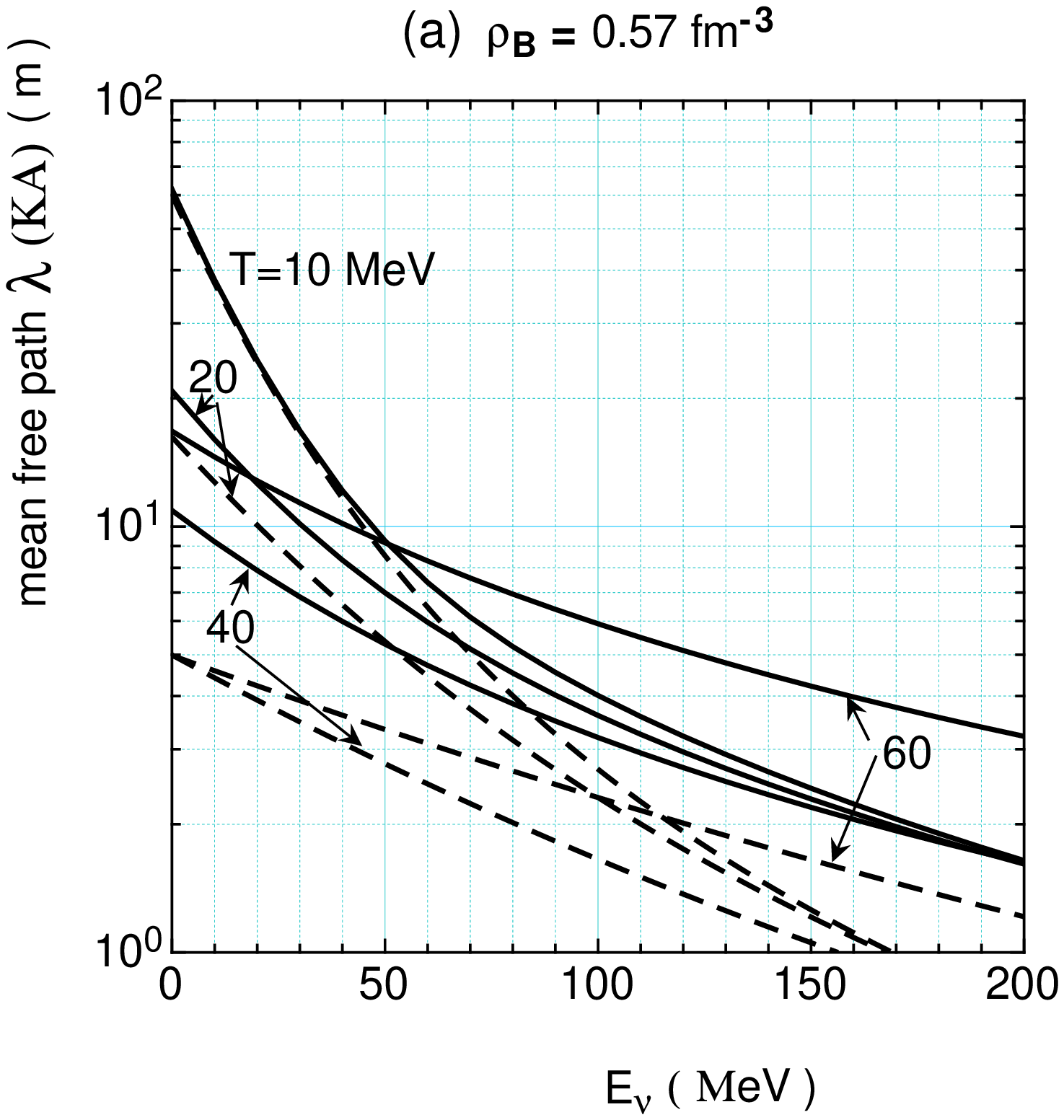}}
\end{minipage}~
\begin{minipage}[r]{0.5\textwidth}
\centerline{
\epsfxsize=\textwidth
\epsffile{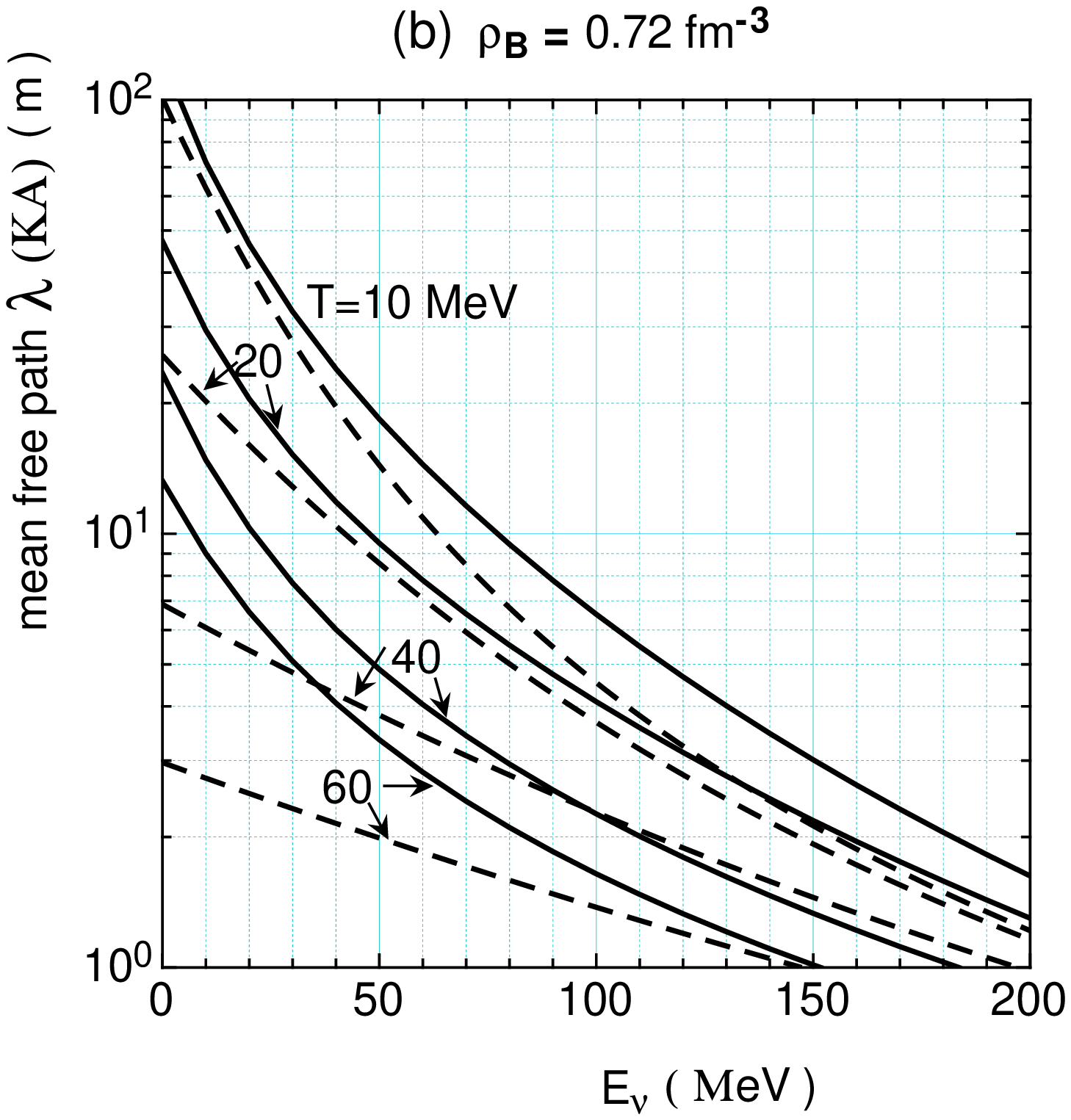}}
\end{minipage}
\caption{(a) The mean free paths $\lambda({\rm KA})$ for KA as functions of the neutrino energy $E_\nu$ for several temperatures ($T$=10, 20, 40, 60  MeV) and the baryon number density $\rhob$=0.57 fm$^{-3}$ in the nondegenerate neutrino case.  
The solid lines denote the results with numerical integration for the phase space integral $I^{(N)}({\rm KA})$ [ Eq.~(39) ] , while the dashed lines denote the results with the analytic formulae Eq.~(41) in the low-temperature approximation. 
(b) The same as Fig. 1~(a) but for the baryon number density $\rhob$=0.72 fm$^{-3}$ . }
\label{fig1}
\end{figure}

In Fig.~\ref{fig1}, we show the mean free paths 
$\lambda$(KA) as functions of the neutrino energy $E_\nu$ for several temperatures ( $T$=10, 20, 40, 60 MeV) and  baryon number densities.  Fig.~\ref{fig1}~(a) is for $\rhob$=0.57 fm$^{-3}$ and Fig.~\ref{fig1}~(b) is for $\rhob$=0.72 fm$^{-3}$. The solid lines denote the results with the numerical integration for the phase space integral $I^{(N)}({\rm KA})$ [ Eq.~(\ref{eq:i2}) ] , while the dashed lines denote the results with the analytic formulae Eq.~(\ref{eq:analytic}) in the low-temperature approximation. 
For temperatures $T\lesssim $10 MeV, where the low-temperature approximations (i)$-$(iii) [see \ref{subsubsec:lowt}] are obviously valid, the analytic results almost agree with the numerically integrated results for the neutrino energies less than the thermally  averaged neutrino energy $\displaystyle \langle E_\nu\rangle\Big(\equiv\frac{7\pi^4}{180\zeta(3)}T\simeq 3.15 T$ with $\zeta(3)$ the zeta function\Big) .  Even at  $T$=60 MeV, one finds 2.7$\lesssim$$\lambda({\rm KA})$(numerical) /$\lambda({\rm KA})$(analytic) $\lesssim$ 3.4 for $\rhob$=0.57 fm$^{-3}$, and 0.97$\lesssim$$\lambda({\rm KA})$(numerical) /$\lambda({\rm KA})$(analytic) $\lesssim$ 4.5 for $\rhob$=0.72 fm$^{-3}$ for $0<E_\nu< 200$ MeV. 
For lower temperatures, the agreement between $\lambda$(KA,numerical) and $\lambda$(KA,analytic) is better. 
Therfore, 
the analytic formulae may be safely applied to see qualitative dependences of the $\lambda$(KA) on the temperature $T$ and neutrino energy $E_\nu$. 

Qualitatively, the approximations (i) $ \widetilde\mu_N\rightarrow \infty$ and (ii) $F_{\rm a}\rightarrow 8\pi^2 \widetilde{E}_\nu $ in the phase space integrals $I^{(N)}({\rm KA})$ make the phase space larger than the actual one, which leads to shorter mean free paths than $\lambda({\rm KA})$(numerical). On the other hand, the approximation (iii) $\widetilde{E}_e\rightarrow \widetilde\mu_e$  restricts the phase space, which  leads to longer mean free paths than $\lambda({\rm KA})$(numerical).  These competing effects determine the relative magnitude of $\lambda({\rm KA})$(analytic) compared with $\lambda({\rm KA})$(numerical). 

The mean free path $\lambda({\rm KA})$ decreases monotonically with the increase in the neutrino energy $E_\nu$ at a fixed temperature. 
 It also decreases with the increase in temperature $T$ at a fixed neutrino energy as far as  the dependence of the parameters on the temperature is weak ($T\lesssim$ 40 MeV). 
 These dependences on $T$ and $E_\nu$ can be understood from the analytic formula for $\lambda$(KA) [Eq.~(\ref{eq:analytic})] . 
 For $T\gtrsim$ 60 MeV and $\rhob$=0.57 fm$^{-3}$, a large part of kaon condensates changes into thermal kaons, and the amplitude of kaon condensation ($\propto \sin^2\theta$) is weakened [see Table~\ref{tab:ndparam}]. As a result, the mean free path $\lambda$(KA) becomes long as compared with that for $T$=40 MeV.  
 
As is expected, $\lambda$(KA) is shorter than that for the ordinary two-nucleon process, $\nu_e nN\rightarrow e^- pN$\cite{ss79,hj87} by several 
orders of magnitude when the temperature $T$ is not very high. 

\subsubsection{Comparison between absorption and scattering processes}
\label{subsubsec:2}

\begin{figure}[tt]\noindent
\begin{minipage}[l]{0.5\textwidth}
\centerline{
\epsfxsize=\textwidth
\epsffile{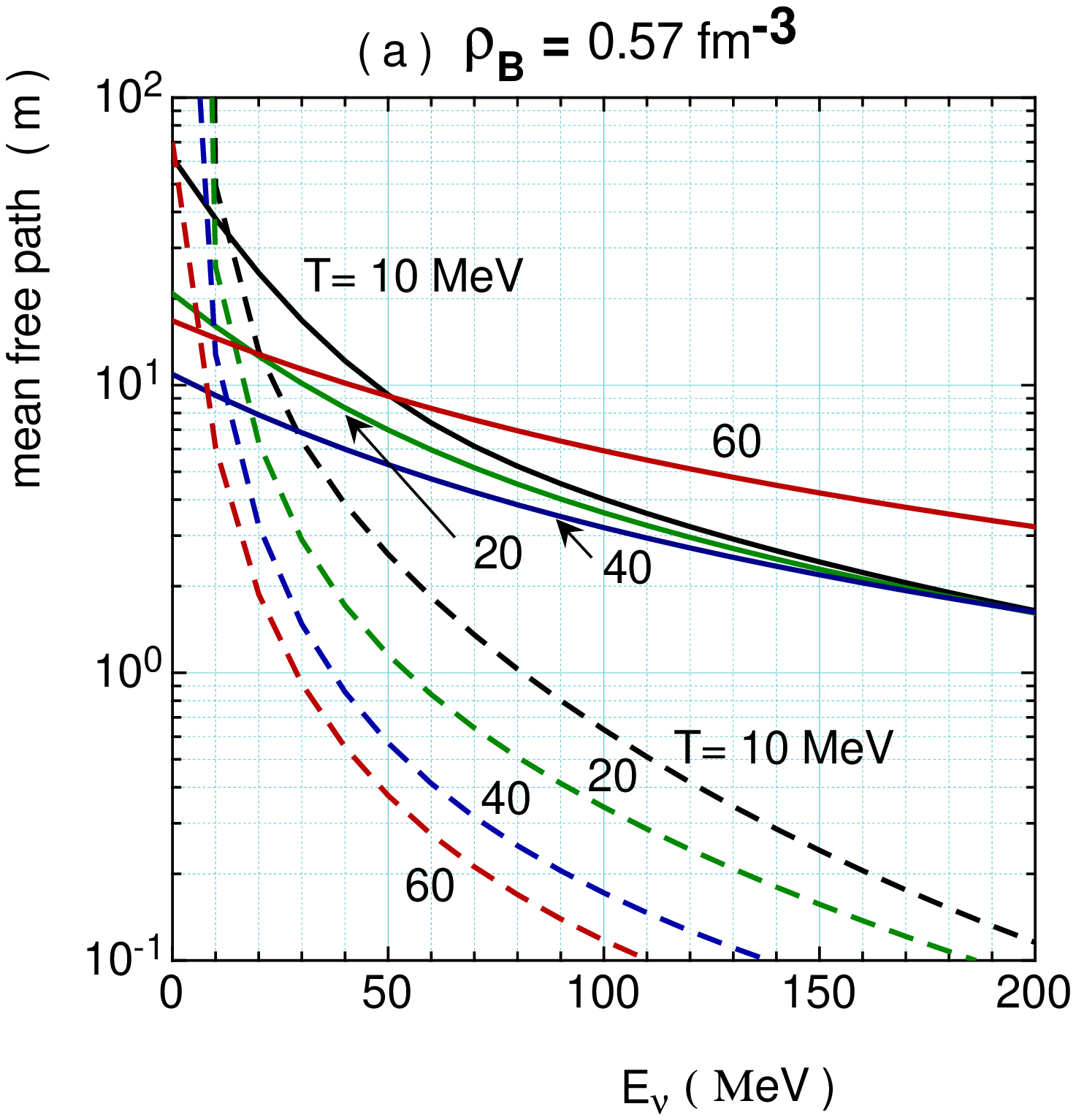}}
\end{minipage}~
\begin{minipage}[r]{0.5\textwidth}
\centerline{
\epsfxsize=\textwidth
\epsffile{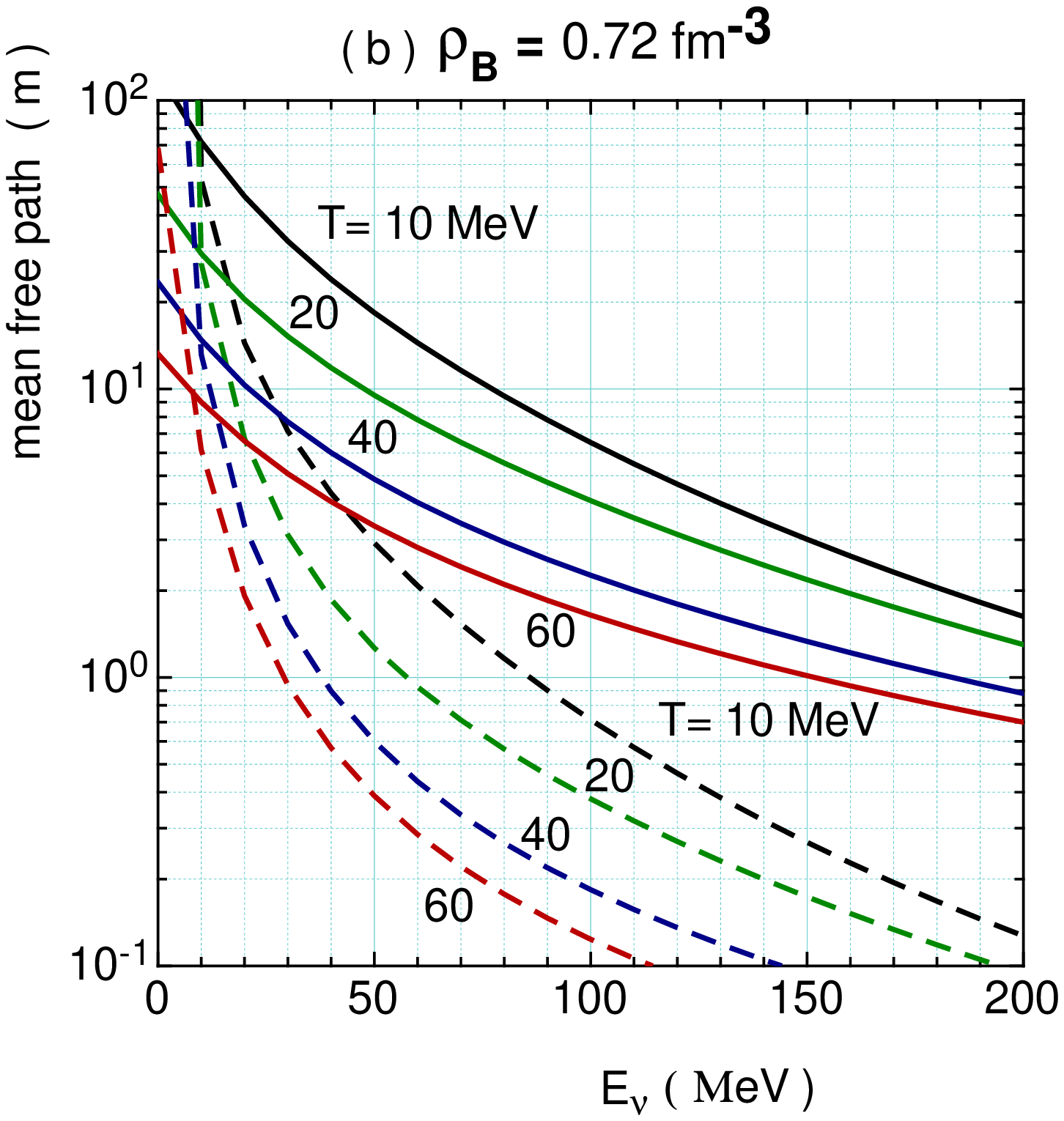}}
\end{minipage}
\caption{(a) The mean free paths $\lambda({\rm KA})$ (solid lines) and $\lambda$(S) (dashed lines)  as functions of  the neutrino energy $E_\nu$ for several temperatures ($T$=10, 20, 40, 60  MeV) and the baryon number density $\rhob$=0.57 fm$^{-3}$ for the  nondegenerate neutrinos.  
(b) The same as Fig. 2~(a) but for the baryon number density $\rhob$=0.72 fm$^{-3}$ . }
\label{fig2}
\end{figure}

Here we compare the mean free paths for KA with those for S [(\ref{eq:scattering})]. All the results have been obtained by performing the phase space integrals numerically.  
In Fig.~\ref{fig2}~(a), the mean free paths $\lambda$(KA) (solid lines) and those $\lambda$(S) for S (dashed lines) are shown as functions of the neutrino energy $E_\nu$ for several temperatures ( $T$=10, 20, 40, 60 MeV) and  baryon number densities $\rhob$.  Fig.~\ref{fig2}~(a) is for $\rhob$=0.57 fm$^{-3}$ and Fig.~\ref{fig2}~(b) is for $\rhob$=0.72 fm$^{-3}$. 

 For $E_\nu \lesssim$ 10 MeV, $\lambda$(S) is much larger than $\lambda$(KA), because the allowable phase space is  restricted much for S. However,  $\lambda$(S) gets smaller than $\lambda$(KA) for  $E_\nu\gtrsim$ 20 MeV. 
At low temperatures, $T\lesssim$ 10 MeV,  the S process is almost forbidden for $E_\nu\sim $ 0. On the other hand, at high temperatures, $T\gtrsim$ 40 MeV, it is operative even at $E_\nu\sim $ 0, because the outgoing neutrino which has enough energy and momentum with $O(T)$ ($\sim$ a few tens of MeV) contributes to the reaction.  In Table~\ref{tab:ndmatrix}, we list the reduced squared matrix elements for KA and S. 
The matrix elements for S are larger than those for KA by one or two orders of magnitude. Therefore, once there is a sufficient phase space for the energetic neutrino, 
the reaction rate for S easily becomes larger than that for KA. 
The difference in the matrix elements between KA and S mainly comes from the following facts: 
 (1)  The KA process is Cabibbo-suppressed by the term  proportional to $\sin^2\theta_{\rm C}$ in Eq.~(\ref{eq:matrixa}). 
 (2) The matrix elements for  KA in Eq.~(\ref{eq:matrixa}) contain a factor proportional to $\sin^2\theta$ which is small for a weak condensate. On the other hand, the S process is free from these suppression factors. 
 
\begin{figure}[tt]\noindent
\centerline{
\epsfxsize=0.5\textwidth
\epsffile{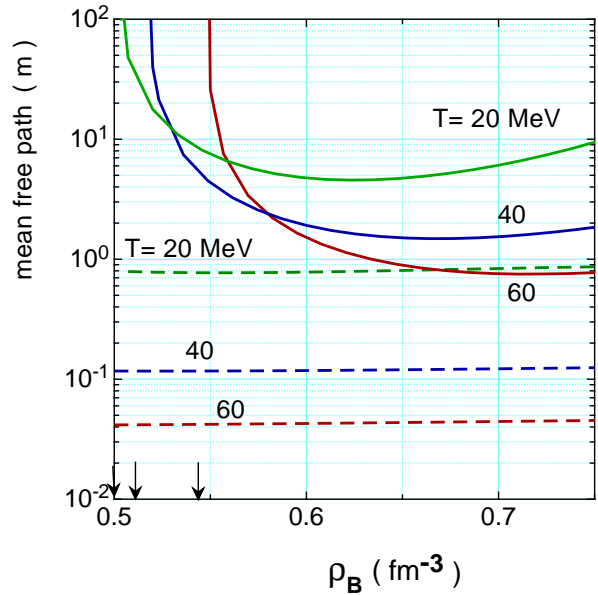}}
\caption{The mean free paths for KA (solid lines) and S (dashed lines) as functions of density $\rhob$ at temperatures $T$=20, 40, 60 MeV for nondegenerate neutrinos. 
The neutrino energy $E_\nu$ is set to be the thermally averaged value ($E_\nu\simeq 3.15T$).
The arrows denote the critical densities for kaon condensation at temperatures $T$=20, 40, 60 MeV. }
\label{fig3}
\end{figure}

 In Fig.~\ref{fig3}, the mean free paths for KA (solid lines) and S (dashed lines) are shown as functions of baryon number density $\rhob$ for several temperatures ($T$=20, 40, 60 MeV). The neutrino energy $E_\nu$ is set equal to the thermally averaged value, $\displaystyle\langle E_\nu\rangle (\simeq 3.15T)$. 
 The arrows denote the critical densities for kaon condensation at temperatures $T$=20, 40, 60 MeV\cite{yt01}. 
 The mean free paths $\lambda$(KA) depend sensitively on density,  
reflecting the combining effects of the chiral angle 
$\theta$ and charge chemical potential $\mu_e(=\mu_K)$ 
which appear in  Eq.~(\ref{eq:analytic}) within the 
low-temperature approximation: 
The chiral angle $\theta$ monotonically increases with increasing 
density as kaon condensation develops, while the kaon 
chemical potential $\mu_K$ decreases with increasing 
density\cite{fmtt94,yt01} and it vanishes at 
$\rhob^\ast\simeq$ 0.80 fm$^{-3}$, beyond which it 
becomes negative\cite{yt01}. As a net effect, 
the matrix elements (\ref{eq:matrixa}) for 
KA increase with density except for density near $\rhob^\ast$, and 
the mean free path $\lambda$(KA) decreases with density. 
Near the typical density $\rhob^\ast$, 
due to the small value of $\mu_K$,  $\lambda$(KA) becomes long and increases 
monotonically with  density.
\footnote{ The mean free path $\lambda$(KA) deverges at $\rhob=\rhob^\ast$. 
  For $\rhob>\rhob^\ast$ ($\mu_K<0$), positrons appear in place of 
electrons, and the KA process (\ref{eq:ab}) is replaced by $\nu_e N \langle K^-\rangle e^+\rightarrow N$. However, the negative value for $\mu_K$ should not be taken seriously, because it has been demonstrated that the monotonous decrease of $\mu_K$ with density $\rhob$ is saturated and $\mu_K$ remains positive over the relevant densities after relativistic corrections are taken into account\cite{mfmt94,fmmt96}.} 
On the other hand, the dependence of the mean free paths $\lambda$(S) on density is small. Over the relevant densities, $\lambda$(KA) is always longer than $\lambda$(S) at a fixed temperature. 

From the above discussions on Figs.~\ref{fig1} and \ref{fig2} , it is concluded that the  KA process has a minor effect on the evolution of protoneutron stars as compared with the S process in most cases of neutrino energies and temperatures for nondegenerate neutrinos.

\section{Degenerate neutrino case}
\label{sec:dege}

\subsection{ Relevant reactions }
\label{subsec:main}

Here we consider neutrino opacities  with kaon condensates in the 
degenerate neutrino case ($\mu_\nu>0$).  We take into account the relevant reactions as follows: (i) the kaon -induced neutrino absorption process (KA) [ (\ref{eq:ab}) ]. (ii) the neutrino-nucleon scattering process (S), and (iii) the direct neutrino absorption process (DA), 
\beq
\nu_e({\bf p}_\nu) +\ n({\bf p}) \rightarrow e^-({\bf p}_e)  + \ p({\bf p}')  \ . 
\label{eq:da}
\eeq

The mean free path for the DA process, $\lambda$(DA), is given as  
\begin{equation}
1/\lambda ({\rm DA})=\frac{G_F^2}{2(2\pi)^5}
\overline{H_{\mu\nu}({\rm DA}) L^{\mu\nu}} I({\rm DA}) \ , 
\label{eq:lda}
\end{equation}
where
\beq
I({\rm DA})=\frac{1}{E_\nu}\int_0^\infty |{\bf p} |d|{\bf p}  |
\int_0^\infty |{\bf p}' |d|{\bf p}' |E_3\Theta(E_3)
f({\bf p})[1-f(E_3)][ 1-f({\bf p}') ]
F(E_\nu,p,E_3,p') \ , 
\label{eq:ida}
\eeq
with
\beq
E_3\equiv \frac{{\bf p}^2}{2m_N}-\frac{{\bf p}'^2}{2m_N}+E_\nu \ . 
\label{eq:e3da}
\eeq
The reduced squared matrix element after angular averaging, $\overline{H_{\mu\nu}({\rm DA}) L^{\mu\nu}}$, is given by 
\beq
\overline{H_{\mu\nu}({\rm DA}) L^{\mu\nu}}
=4(1+3g_A^2)\cos^2\theta_{\rm C}\cos^2\frac{\theta}{2} \ . 
\label{eq:matrixda}
\eeq
It is to be noted that the expression is the same as that for the direct URCA process, $n\rightarrow pe^-\bar\nu_e$, $pe^-\rightarrow n \nu_e$,  which is relevant to the neutron star cooling.  
For details of the derivation of (\ref{eq:matrixda}), See Ref.~\cite{fmtt94}.  

The expressions of the mean free paths for KA and S [ (\ref{eq:la}) and (\ref{eq:finals}), respectively] are the same as those for  nondegenerate neutrinos. For KA, 
however, it is to be noted that the statistical factor for the electron, $\displaystyle 1/[\exp\lbrace -(x_2+x_4+\widetilde E_\nu)\rbrace +1]$,  in the reduced expression for the phase-space integral $I^{(N)}$(KA) [ (\ref{eq:i2}) ]  is to be replaced by 
$\displaystyle 1/[\exp\lbrace -(x_2+x_4+\widetilde E_\nu-\widetilde\mu_\nu)\rbrace +1]$ 
with $\widetilde\mu_\nu=\mu_\nu/T$, which results from the chemical equilibrium relation, $\mu_e-\mu_K=\mu_\nu$ in the degenerate neutrino case. As a result, the analytic formula $\lambda^{(N)}$(KA) in the low-temperature approximation is also modified from that for nondegenerate neutrinos by replacement $\widetilde E_\nu\rightarrow \widetilde E_\nu-\widetilde\mu_\nu$ in Eqs.~(\ref{eq:opac-pn}) and (\ref{eq:analytic}), and written as 
\beq
1/\lambda^{(N)}({\rm KA})(E_\nu, T)_{{\rm degenerate} \ \nu}=\frac{G_F^2}{16\pi^3}
\overline{H_{\mu\nu}^{(N)}({\rm KA}) L^{\mu\nu}} \cdot
m_N^2 \mu_e T^2\frac{\pi^2+(\widetilde{E}_\nu-\widetilde\mu_\nu)^2}{1+e^{-(\widetilde{E}_\nu-\widetilde\mu_\nu)}} \ . 
\label{eq:analyticadege} 
\eeq
for degenerate neutrinos. 

\subsection{Numerical results in the degenerate neutrino case }
\label{subsec:resultdege}
 
\begin{figure}[t]
\centerline{
\epsfxsize=0.5\textwidth\epsffile{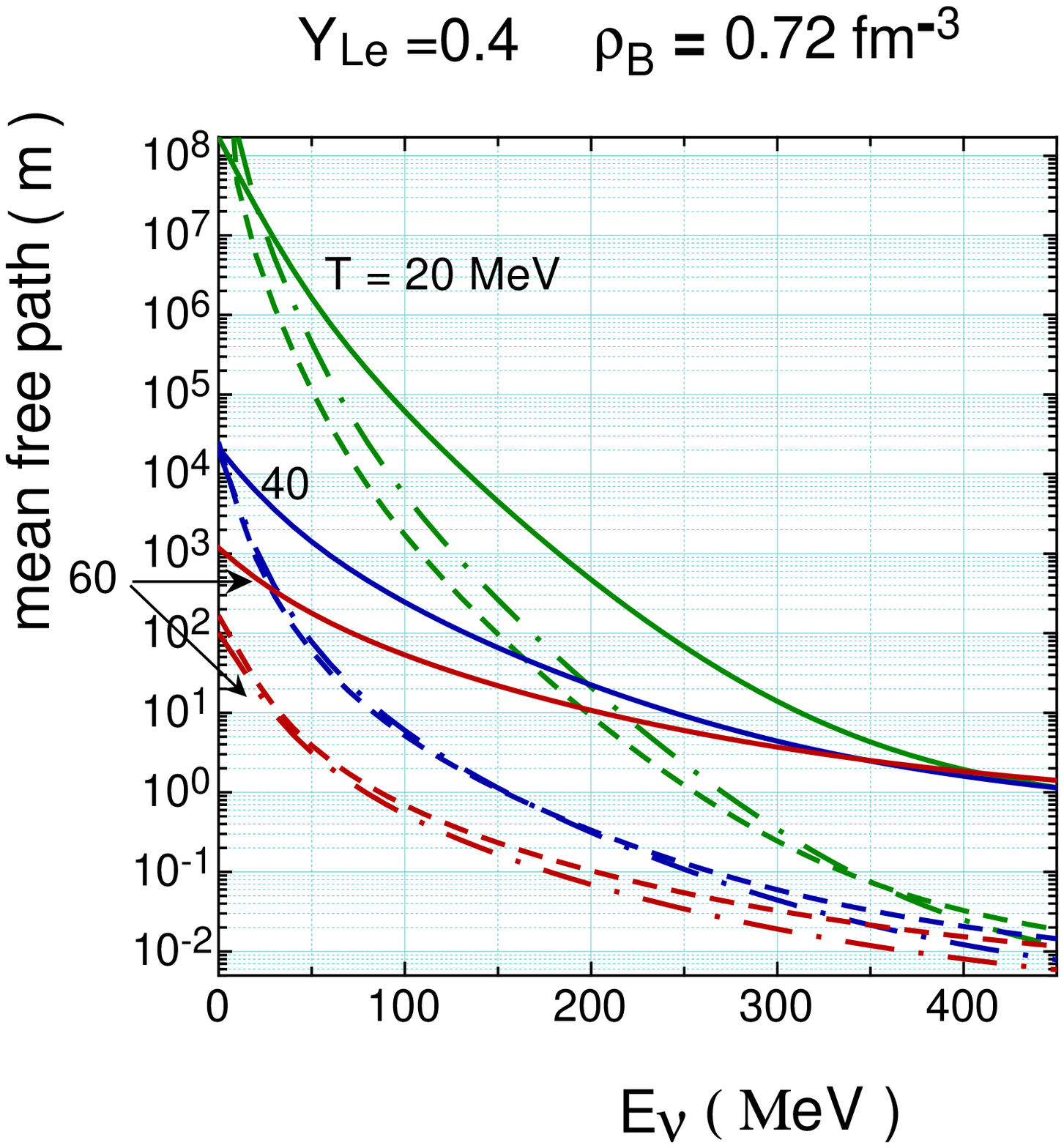}}
\caption{The mean free paths for KA (solid lines), DA (dash-dotted lines) and for S (dashed lines) as functions of  the neutrino energy $E_\nu$ for $\rhob$=0.72 fm$^{-3}$, the lepton fraction $Y_{le}$=0.4, and $T$=20, 40, 60 MeV. for degenerate neutrinos.  }
\label{fig4}
\end{figure}

In Fig.~\ref{fig4}, the mean free paths for KA (solid lines), DA (dash-dotted lines),  and S (dashed lines) are shown as functions of the neutrino energy $E_\nu$ for $\rhob$=0.72 fm$^{-3}$, the lepton fraction $Y_{le}$=0.4, and $T$=20, 40, 60 MeV. All the results are obtained from the numerical integration for the phase space. 
In Table~\ref{tab:dparam}, the physical parameters for the EOS with kaon condensates for  several temperatures $T$ and $\rhob$=0.72 fm$^{-3}$ in the 
neutrino-trapping case with $Y_{le}$=0.4 are listed. 
It is to be noted that, for $Y_{le}$=0.4, the critical density for kaon condensation is larger than 0.62 fm$^{-3}$ [see Ref.~\cite{yt01}], so that we show only the $\rhob$=0.72 fm$^{-3}$ case for the mean free paths in the presence of kaon condensation. 

For most of the incident neutrino energy $E_\nu$, DA and S have dominant contributions to the opacities as compared with KA. 
 At a typical energy $E_\nu=\mu_\nu$ ($\simeq$ 300 MeV), 
$ \lambda$(KA)=10 [m] $\rightarrow$ 4 [m], while $ \lambda$(DA)$\sim \lambda$(S)=0.2 [m] $\rightarrow$ 0.03 [m] as $T$=20 MeV$\rightarrow$ 60 MeV. In Table~\ref{tab:dmatrix}, the reduced matrix elements for KA, DA, and S [
Eqs.~(\ref{eq:matrixa}), (\ref{eq:matrixda}), and (\ref{eq:matrixs}), respectively] for several temperatures $T$ and $\rhob$=0.72 fm$^{-3}$ in the degenerate neutrino case are listed. 
The matrix element for KA is smaller than those for S and DA by one or two orders of magnitude, which stems from both effects of the 
Cabibbo-suppression and the different $\theta$-dependence of the matrix element for KA as compared with S and DA, just as was seen in the nondegenerate neutrino case. This difference in the matrix elements mainly causes the difference of the mean free paths. 
  
Next we compare the mean free path for  KA with that for nondegenerate neutrinos [see Figs.~\ref{fig2}~(b) and \ref{fig4}] . For the incoming neutrino energy $E_\nu$ being far below the chemical potential of neutrino (i. e., $E_\nu \ll \mu_\nu=\mu_e-\mu_K <\mu_e\sim$350 MeV),  the difference of energy or  momenum between the incoming and outgoing degenerate leptons is large in the degenerate neutrino case. This large difference leads to kinematical restriction of the allowable phase space for KA. Thus the mean free paths for KA for degenerate neutrinos are much larger than those for nondegenerate neutrinos at given $E_\nu$ and $T$. The significant reduction of the reaction rate for KA is also seen from the exponentially damping factor $\exp(\widetilde\mu_\nu)$ in the analytic expression Eq.~(\ref{eq:analyticadege}) of the mean free path for KA in the  degenerate neutrino case.  For a large $E_\nu (\sim \mu_\nu$), the mean free paths for KA for both degenerate and nondegenerate neutrinos are the same order of magnitude. 
 
Also for the S process, the allowable phase space is much restricted for the incoming neutrino energy $E_\nu\ll \mu_\nu$, because the outgoing neutrino is degenerate with $\mu_\nu\sim$300 MeV. Thus the mean free paths for S for degenerate neutrinos with $E_\nu\ll \mu_\nu$ are much larger than those for nondegenerate neutrinos at given $E_\nu$ and $T$, while for a large $E_\nu (\sim \mu_\nu$), the mean free paths for S for both degenerate and nondegenerate neutrinos are the same order of magnitude.
 
\section{Cooling time scale by neutrino diffusion}
\label{sec:cooling}

We estimate a cooling time scale in the nondegenerate neutrino case driven by neutrino diffusion taking into account both the KA and S processes. We take a simple neutron star model having a kaon-condensed core with uniform density $\rhob$, temperature $T$ and a radius $r$. The cooling time scale for neutrino diffusion is roughly estimated from Eq.~(\ref{eq:dtdt}) as
\beq
\tau=(\lambda^R)^{-1} \ r^2 \ C_{\rm V} \ \sigma_{\rm SB}^{-1} \ T^{-3} \ , 
\label{eq:tau}
\eeq
where the Rosseland mean free path $\lambda^R$ is given by 
\beq
\lambda^R=\frac{\int dE_\nu E_\nu^3\Bigg\lbrack 
\lambda({\rm KA})^{-1}(1+e^{-E_\nu/T})+\lambda({\rm S})^{-1}\Bigg\rbrack^{-1}\frac{d}{dT}(1+e^{E_\nu/T})^{-1}}{\int dE_\nu E_\nu^3\frac{d}{dT}
(1+e^{E_\nu/T})^{-1}} \ . 
\label{eq:lr}
\eeq
For simplicity, the free Fermi gas model is used for estimation of $C_{\rm V}$. The time scale $\tau$ is listed in Table~\ref{tab:tau} for $\rhob$=0.57 fm$^{-3}$ and $r$=5 km. One has several tens of second for $\tau$ for $T$=10$-$40 MeV. As is discussed in Sec.~\ref{subsubsec:2}, the main contribution to the opacities comes from the S process, and the KA process is negligible. If we take into account nuclear correlation effects on the opacities which have been recently considered by several authors\cite{rpl98,rplp99,nhv99,nb01,y00}, the inverse of the Rosseland mean opacity would become large and thus the cooling time $\tau$ would be reduced as compared with those without nuclear correlation effects. 

\section{Summary and concluding remarks}
\label{sec:summary}

On the basis of chiral symmetry, we have obtained the neutrino mean free paths relevant to protoneutron stars with kaon condensates in both nondegenerate and degenerate neutrino cases. 
The matrix elements have been obtained by the use of current algebra, and mean free paths have been numerically calculated with the help of the parameters consistently obtained with the kaon-condensed EOS at finite temperatures.   
The mean free path for the kaon-induced neutrino absorption   process (KA) has been compared with that for the neutrino-nucleon scattering process (S) and 
the direct neutrino absorption process (DA). It has been shown that the mean free path for KA is longer than those for S and DA by roughly one or two orders of magnitude at fixed temperature, density and neutrino energy in both nondegenerate and degenerate neutrino cases. Thus it is concluded that KA is not important as compared with DA and S for dynamical and thermal evolution of protoneutron stars in both the deleptonization stage where neutrinos are degenerate 
and the initial cooling stage where neutrinos are nondegenerate. 

Recently, it has been suggested that coherent scattering off a droplet with kaon condensates in a mixed phase may have a dominant contribution to the neutrino opacity\cite{rbp00}. This process can occur in the presence of droplets as a result of a first order phase transition\cite{g01}. However, it has also been shown that stability of a mixed phase largely depends on charge screening and surface effects so that these effects should be carefully taken into account for the discussion of the presence of the mixed 
phase\cite{hps93,c00,nr01,n02,vty01}. 

For future work, the neutrino opacities can be applied to a dynamical simulation of a delayed collapse scenario of a protoneutron star accompanying a phase transition to a kaon-condensed phase with consistently taking into account  modification of the EOS during the deleptonization and subsequent initial cooling era. 

\section{Acknowledgements}
\label{sec:ack}

 This work is supported in part by the Japanese Grant-in-Aid for 
Scientific Research Fund (C) of the Ministry of Education, Science, 
Sports, and Culture (Nos. 12640289, 13640282, 13640291) and the National Science Foundation under Grant Nos. PHY-9722138 and PHY-0071006. Numerical calculations were carried out on the DEC Alpha Server 4100 System, Chiba Institute of Technology. 

\appendix
\section{Phase space integrals for KA}
\label{sec:phase}

The phase space integral $I^{(N)}$(KA) for KA in Eq.~(\ref{eq:la}) is given by 
\beq
I^{(N)}({\rm KA})=\int_0^\infty \magpi^2 d\magpi \int_0^\infty \magpe^2 d\magpe 
\int_0^\infty \magpf^2 d\magpf
S_a
\delta(E_e+E'_{N}-E_\nu-E_N-\mu_K)\cdot A 
\label{eq:i}
\eeq
with $A$ being the angular integral:
\beqa
A&=&\int d\Omega_N \int d\Omega_e \int d\Omega_{N'}
\delta^{(3)}({\bf p_e}+{\bf p}_{N}'-{\bf p}_\nu-{\bf p}_N) \cr
&=&\frac{32\pi}{\magpnu\magpi\magpe\magpf}\int_0^\infty d x\frac{\sin(\magpnu x)\sin(\magpi x)\sin(\magpe x)\sin(\magpf x)}{x^2} \cr
&=&F_a(\magpnu , \magpi , \magpe , \magpf)/(\magpnu\magpi\magpe\magpf) \ , 
\label{eq:angular}
\eeqa
The function $F_a$ may be written as 
$F_a(\magpnu , \magpi , \magpe , \magpf)=-16\pi (F_1-F_2)$ with
\beqa
F_1\equiv \cases{\pi \ {\rm min}(\magpnu,\magpi)  
\qquad \textrm{for} \quad 0<\magpe+\magpf <|\magpnu-\magpi| \cr
                                \pi(\magpnu+\magpi-\magpe-\magpf)/4 
\qquad \textrm{for} \quad |\magpnu-\magpi| <\magpe+\magpf <\magpnu+\magpi\cr
                                0
\qquad \textrm{for} \quad \magpnu+\magpi<\magpe+\magpf } 
\label{eq:f1}
\eeqa
and 
\beqa
F_2\equiv \cases{\pi \ {\rm min}(\magpnu,\magpi)  
\qquad \textrm{for} \quad 0<|\magpe-\magpf|<|\magpnu-\magpi| \cr
                                \pi(\magpnu+\magpi-|\magpe-\magpf|)/4 
\qquad \textrm{for} \quad |\magpnu-\magpi| <|\magpe-\magpf| <\magpnu+\magpi\cr
                                0
\qquad \textrm{for} \quad \magpnu+\magpi<|\magpe-\magpf| \ .}   
\label{eq:f2}
\eeqa

\begin{table}
\begin{tabular}{c c || c c c c c}
$\rhob({\rm fm}^{-3}) $  & $T$(MeV) &
$\theta$(rad) &  $\mu_p^0$(MeV) &  $\mu_n^0$(MeV) &   $\mu_K$(MeV)   & $\mu_e$(MeV) \cr\hline
0.57    &10 & 0.67  & 62  &    105   &   164   &   164 \cr
           & 20& 0.64  & 58  &    102   &   163   &   163  \cr
          & 40 & 0.58  & 39  &      89   &   158   &   158 \cr
          & 60 & 0.37  & 7  &      69   &   157   &   157 \cr
          \hline
0.72  & 10   & 1.16  & 101&      98    &    45   &     45 \cr
       &  20   & 1.15  &   98 &     95   &     44   &     44  \cr
       &  40   & 1.13  &   86 &     83   &     42  &      42 \cr
       &  60   & 1.08  &   63 &     63   &     46  &      46 \cr
\end{tabular}
\smallskip
\caption{Physical parameters for the EOS with kaon condensates for several temperatures $T$  and baryon number densities $\rho_{\rm B}$ without trapped neutrinos (the electron lepton fraction $Y_{le}=0$)[15]. }
\label{tab:ndparam} 
\end{table}

\begin{table}
\begin{tabular}{cc||c|c|c|c}
 & & \multicolumn{2}{c|}{KA}&\multicolumn{2}{c}{S} \\\hline
$\rhob$({\rm fm}$^{-3}$) & $T$ (MeV) &
\ \ $\overline{H_{\mu\nu}^{(p)}({\rm KA})L^{\mu\nu}}$ \ \ & 
\ \ $\overline{H_{\mu\nu}^{(n)}({\rm KA})L^{\mu\nu}}$ \ \ & 
\ \ $\overline{H_{\mu\nu}^{(p)}({\rm S}) L_Z^{\mu\nu}}$ \ \ & 
\ \ $\overline{H_{\mu\nu}^{(n)}({\rm S}) L_Z^{\mu\nu}}$ \ \ \\
\hline\hline 
 0.57 & 20 & 0.13 & 0.029 & 4.1 & 5.4 \\
         & 40 & 0.11 & 0.024 & 4.2 & 5.5 \\
         & 60 & 0.048 & 0.011 & 4.5 & 5.6 \\
\hline 
0.72 & 20 & 0.30  & 0.067  & 2.9 & 5.0 \\
	& 40 & 0.29 & 0.065 & 3.0 & 5.0 \\
	& 60 & 0.28 & 0.062 & 3.1 & 5.0 \\ 
\end{tabular}
\smallskip
\caption{The reduced squared matrix elements for KA and S for several temperatures $T$ and  baryon number densities $\rhob$ for nondegenerate neutrinos. }
\label{tab:ndmatrix}
\end{table}

\begin{table}
\begin{tabular}{c||c|c|c|c|c|c}
$T$ (MeV) & $\theta$ (rad) & $\mu_p^0$ (MeV) & $\mu_n^0$ (MeV) & 
$\mu_K$ (MeV) & $\mu_e$ (MeV) & $\mu_\nu$ (MeV) \\
\hline\hline
 20 & 0.60 & 92 & 102 & 48 & 364 & 315 \\
 \hline
 40 & 0.53 & 78 & 91 & 52 & 353 & 301 \\
 \hline
 60 & 0.42 & 55 & 71 & 51 & 334 & 283 \\
\end{tabular}
\smallskip
\caption{Physical parameters for the EOS with kaon condensates for several temperatures $T$ and $\rhob$=0.72 fm$^{-3}$ in the neutrino-trapping case ($Y_{le}=0.4$)[15]. }
\label{tab:dparam}
\end{table}
 
\begin{table}
\begin{tabular}{c||c|c|c|c|c}
 &\multicolumn{2}{c|}{KA}&\multicolumn{1}{c|}{DA}&\multicolumn{2}{c}{S} \\\hline
$T$ (MeV) &
\ \ $\overline{H^{(p)}_{\mu\nu}({\rm KA}) L^{\mu\nu}}$ \ \ & 
\ \ $\overline{H^{(n)}_{\mu\nu}({\rm KA}) L^{\mu\nu}}$ \ \ & 
\ \ $\overline{H_{\mu\nu}({\rm DA}) L^{\mu\nu}}$ \ \ & 
\ \ $\overline{H^{(p)}_{\mu\nu}({\rm S}) L_{\rm Z}^{\mu\nu}}$ \ \ & 
\ \ $\overline{H^{(n)}_{\mu\nu}({\rm S}) L_{\rm Z}^{\mu\nu}}$ \ \ \\
\hline
20 &  0.11  & 0.026 &20  & 4.1 & 5.5 \\
\hline
40 & 0.090 & 0.020 & 20 & 4.3 & 5.5 \\
\hline
60 & 0.060 & 0.013 & 21 & 4.4 &5.6 \\
\end{tabular} 
\smallskip
\caption{The reduced squared matrix elements for KA, DA, and S for several temperatures $T$ and $\rhob$=0.72 fm$^{-3}$ in a degenerate neutrino case ($\mu_\nu>0$). }
\label{tab:dmatrix}
\end{table}

\begin{table}
\begin{tabular}{c||c|c}
$T$ (MeV) & $\lambda^R$ (m) & $\tau$ (sec) \\
\hline\hline
10 & 2.5 & 33 \\
\hline
40 & 0.07 & 70 \\
\end{tabular}
\smallskip
\caption{The Rosseland mean free paths $\lambda^{\rm R}$ [Eq.~(67) ] and the initial cooling time scales $\tau$ by neutrino diffusion [Eq.~(66) ]  estimated at several temperatures $T$ for a simplified neutron star model. }
\label{tab:tau}
\end{table}

\newpage

\begin{thebibliography}{99}

\bibitem{kn86} D.~B.~Kaplan and A.~E.~Nelson, Phys.~Lett.~{\bf B175}, 
57 (1986); {\bf B179}, 409(E) (1986). \\
 A.~E.~Nelson and D.~B.~Kaplan, Phys.~Lett.~{\bf B192},  
193 (1987). 
\bibitem{t95} T.~Tatsumi, Prog.~Theor.~Phys.Suppl. {\bf 120},111
(1995). 
\bibitem{l96}  C.~-H.~Lee, Phys.~Rep.{\bf 275},197 (1996). 
\bibitem{pbpelk97} M.~Prakash, I.~Bombaci, M.~Prakash, P.~J.~Ellis,
J.~M.~Lattimer, R.~Knorren, Phys. Rep. {\bf 280}, 1(1997).
\bibitem{bb94} G.~E.~Brown and H.~A.~Bethe, 
Astrophys.J.{\bf 423}, 659 (1994).
\bibitem{bst96} T.~W.~Baumgarte, S.~L.~Shapiro and S.~Teukolsky,
Astrophys. J. {\bf 443}, 717 (1995); {\bf 458}, 680 (1996).
\bibitem{tt88} T.~Takatsuka, Prog.~Theor.~Phys.~{\bf 80}, 361 (1988);
{\it ibid} {\bf 82},475 (1989); \\ 
in {\it The Structure and Evolution of Neutron Stars, 
Kyoto,1990}, edited by D.~Pines, R.~Tamagaki, and S.~Tsuruta (Addison-Wesley,
California,1992), p.257.
\bibitem{ts88} M.~Takahara and K.~Sato, Prog.~Theor.~Phys.~{\bf 80},861 (1988). 
\bibitem{g95} N.~K.~Glendenning, Astrophys.~J.~{\bf 448}, 797 (1995). 
\bibitem{kj95} W.~Keil and H.-Th.~Janka, Astron.~Astrophys.~{\bf 296},145 (1995).
\bibitem{p99} J.~A.~Pons,  et al, Astrophys.~J.~{\bf 513}, 780 (1999). 
\bibitem{bh89} O.~G.~Benvenuto and J.~E.~Horvath, Phys.~Rev.~Lett.~{\bf
63}, 716 (1989). 
\bibitem{pcl95} M.~Prakash, J.~Cooke and J.~M.~Lattimer, 
Phys.~Rev.~{\bf D52}, 661 (1995). 
\bibitem{ty98} T.~Tatsumi and M.~Yasuhira, Phys.~Lett.~{\bf B 441}, 9 (1998); Nucl.~Phys.~{\bf A 653}, 133 (1999); Nucl. Phys. ~{\bf A670}  218c (2000).
\bibitem{yt01} M.~Yasuhira and T.~Tatsumi, Nucl.~Phys.~{\bf A 663$\&$664}, 881c (2000); 
Nucl.~Phys.~{\bf A 680}, 102c (2001); 
Nucl.~Phys.~{\bf A 690}, 769 (2001). \\
T.~Tatsumi and M.~Yasuhira, Proc. of INPC 2001 edited by 
E. Norman et al, American Inst. of Phys, 476 (2002).  
\bibitem{p00} J.~A.~Pons, S.~Reddy, P.~J.~Ellis, M.~Prakash and J.~M.~Lattimer, Phys.~Rev.~{\bf C 62}, 035803 (2000).
\bibitem{p01} J.~A.~Pons, J.~A.~Miralles, M.~Prakash and J.~M.~Lattimer, Astrophys.~J.~{\bf 553}, 382 (2001). 
\bibitem{mti00} T.~Muto, T.~Tatsumi, and N.~Iwamoto, Phys.~Rev.~{D 61}, 063001 (2000) ; 
Phys.~Rev.~{D 61}, 083002 (2000). 
\bibitem{pls01} For a review article, M.~Prakash, J.~M.~Lattimer, R.~F.~Sawyer, and R.~R.~Volkas, Ann.~Rev.~Nucl.~Part.~Sci.~{\bf 51}, 295 (2001). 
\bibitem{rpl98} S.~Reddy, M.~Prakash, and J.~M.~Lattimer, 
Phys.~Rev.~{\bf D 58}, 013009 (1998). 
\bibitem{ss79} R.~F.~Sawyer and A.~Soni, Astrophys.~J.~{\bf 230}, 859 (1979). 
\bibitem{lpph91} J.~M.~Lattimer, C.~J.~Pethick, M.~Prakash, and P.~Haensel, 
Phys.~Rev.~Lett.{\bf 66}, 2701(1991). 
\bibitem{fmtt94} H.~Fujii, T.~Muto, T.~Tatsumi and R.~Tamagaki, 
Nucl.Phys.~{\bf A 571},758 (1994); Phys.~Rev.~{\bf C 50}, 3140 (1994). 
\bibitem{fm79} B.~L.~Friman and O.~V.~Maxwell, 
Astrophys.J.~{\bf 232}, 541(1979).
\bibitem{hj87} P.~Haensel and A.~J.~Jerzak, Astron.~Astrophys. {\bf 179}, 127 (1987). 
\bibitem{rplp99} S.~Reddy, M.~Prakash, J.~M.~Lattimer, and J.~A.~Pons, Phys.~Rev.~{\bf C 59}, 2888 (1999). 
\bibitem{ip82} N.~Iwamoto and C.~J.~Pethick, Phys.~Rev.~{\bf D 25}, 313 (1982). 
\bibitem{m93} T.~Muto, Prog.~Theor.~Phys.~{\bf 89}, 415 (1993).
\bibitem{kvk95} E.~E.~Kolomeitsev, D.~N.~Voskresensky, and B.~K\"ampfer, Nucl.~Phys.~{\bf A 588}, 889 (1995). 
\bibitem{m02} T.~Muto, Nucl.~Phys.~{\bf A 691}, 447c (2001); Nucl.~Phys.~{\bf A 697}, 225 (2002).
\bibitem{te97}
V. Thorsson and P.J. Ellis,
Phys. Rev. {\bf D55} (1997) 5177.
\bibitem{ains}
M. Prakash, T.L. Ainsworth and J.M. Lattimer,
Phys. Rev. Lett. {\bf 61}, 2518 (1988).
\bibitem{b88}
G.~E.~Brown, K.~Kubodera, D.~Page and P.~Pizzochero, 
Phys.~Rev.~{\bf D 37}, 2042 (1988). 
\bibitem{t88} T.~Tatsumi, Prog.~Theor.~Phys.~{\bf 80}, 22 (1988). 
\bibitem{jm98} C.~-R.~Ji and D.~-P.~Min, 
Phys.~Rev.~{\bf D57}, 5963 (1998). 
\bibitem{mfmt94} T.~Maruyama, H.~Fujii, T.~Muto and T.~Tatsumi,
Phys.~Lett.~{\bf B 337}, 19 (1994)
\bibitem{fmmt96} H.~Fujii, T.~Maruyama, T.~Muto and T.~Tatsumi,
Nucl.~Phys.{\bf A597}, 645 (1996).
\bibitem{nhv99} J.~Navarro, E.~S.~Hern{\' a}ndez and D.~Vautherin, 
Phys.~Rev.~{\bf C 60}, 045801 (1999). 
\bibitem{nb01} R.~Niembro, P.~Bernardos, M.~L{\' o}pez-Quelle and S.~Marcos, Phys.~Rev.~{\bf C 64}, 055802 (2001). 
\bibitem{y00} S.~Yamada, Nucl.~Phys.~{\bf A 662}, 219 (2000). 
\bibitem{rbp00} S.~Reddy, G.~Bertsch and M.~Prakash, Phys.~Lett.~{\bf B 475}, 1 (2000). 
\bibitem{g01} N.~K.~Glendenning, Phys.~Rep.~{\bf 342}, 393 (2001). 
\bibitem{hps93} H.~Heiselberg, C.J. ~Pethick and E.F. Staubo, Phys. Rev. 
Lett. {\bf 70}, 1355 (1993).   
\bibitem{c00} M.~Christiansen, N.~K.~Glendenning and J.~Schaffner-Bielich, Phys.~Rev.~{\bf C 62}, 025804 (2000). 
\bibitem{nr01} T.~Norsen and S.~Reddy, Phys.~Rev.~{\bf C 63}, 065804 (2001). 
\bibitem{n02} T.~Norsen, Phys.~Rev.~{\bf C 65}, 045805 (2002).
\bibitem{vty01} D.~N.~Voskresensky, M.~Yasuhira and T.~Tatsumi,  Phys. Lett.
{\bf B 541},  93(2002); Nucl. Phys. {\bf A}, submitted (nucl-th/0208067).\\
T. ~Tatsumi, M.~Yasuhira and D.~N.~Voskresensky, Proc. of 7th Int. Sympo. of ``Nuclei in 
the Cosmos'', Nucl. Phys. {\bf A}, in press (nucl-th/0209091). 
\end{thebibliography}
\end{document}